\documentclass[11pt]{article}
\usepackage{jheppub}
\usepackage{graphicx}
\usepackage{amsfonts,amssymb,amsmath}
\usepackage[hang]{subfigure}
\usepackage[a4paper,top=1.89in, bottom=0in, left=2in, right=0in]{geometry}

\newcommand{\be}{\begin{equation}}
\newcommand{\ee}{\end{equation}}
\newcommand{\ba}{\begin{eqnarray}}
\newcommand{\ea}{\end{eqnarray}}

\title{Critical Phenomena of Born-Infeld-de Sitter Black Holes in Cavities}
\author{Fil Simovic$^{1,2}$, Robert B. Mann$^2$}
\affiliation[1]{Perimeter Institute for Theoretical Physics,\\
	31 Caroline St. N.,Waterloo, Ontario N2L 2Y5, Canada}
\affiliation[2]{Department of Physics and Astronomy, University of Waterloo,\\
	Waterloo, Ontario N2L 3G1, Canada}
\emailAdd{fil.simovic@gmail.com, rbmann@uwaterloo.ca}
\abstract{ We examine the thermodynamic behaviour of charged, asymptotically de Sitter black holes embedded in a finite-radius isothermal cavity, with a Born-Infeld gauge field replacing the ordinary Maxwell field. We find that the non-linearities of Born-Infeld theory lead to the presence of reentrant phase transitions in the canonical ensemble, whose existence and character are determined by the maximal electric field strength of the theory. We also examine the phase structure in the grand canonical ensemble, and demonstrate the presence of a new reentrant phase transition from radiation, to an intermediate size black hole, and back to radiation.}

\begin{document}
\maketitle

\section{Introduction}

The study of asymptotically de Sitter black holes is relevant not only for understanding the properties of astrophysical black holes, but also in light of the recently formulated de Sitter/conformal field theory correspondence (dS/CFT) \cite{strominger_ds/cft_2001,balasubramanian_mass_2002,ghezelbash_action_2002}.   These black holes  present unique challenges that their anti-de Sitter (AdS) counterparts do not possess. For one, there is no Killing vector that is timelike everywhere outside the black hole horizon, preventing one from having a well defined asymptotic mass \cite{bousso_adventures_2002,clarkson_eguchi-hanson_2006,clarkson_entropic_2003,ashtekar_asymptotics_2015}.   The notion of a vacuum state is also problematic since the spacetime is essentially nonstationary  \cite{mottola_particle_1985,allen_vacuum_1985,hartong_problems_nodate}, and there  is no natural analog of the Bondi news to characterize gravitational radiation in the full non-linear context \cite{ashtekar_asymptotics_2015}.

These considerations motivate a study of black hole thermodynamics in asymptotically de Sitter spacetimes \cite{sekiwa_thermodynamics_2006,dolan_compressibility_2014,kubiznak_thermodynamics_2016}.  New issues arise due to 
 the presence of a cosmological horizon, and the subject is presently not well understood.  While a temperature can be ascribed to this horizon  \cite{gibbons_cosmological_1977}, it is not in general the same
 as the temperature of the black hole horizon, preventing  the assignment of a single equilibrium temperature to the entire spacetime.  Various proposals have been employed to address this issue,
 including the adoption of an {\it effective temperature}   \cite{urano_mechanical_2009} or by considering each horizon as a separate thermodynamic system  \cite{sekiwa_thermodynamics_2006}.  Another possibility is to add scalar hair to the black hole in a manner that ensures equilibrium \cite{mbarek_reverse_2019}; in  this case a 
 `reverse Hawking-Page transition' is observed, in which the small black hole remains stable as it radiates (the cosmological horizon restoring the particle flux) unlike its well-known AdS counterpart
 \cite{hawking_thermodynamics_1983}.

Another way of understanding the issue is that asymptotically de Sitter black hole lack a natural confining `box' with which to achieve thermodynamic equilibrium. In anti-de Sitter space, this is accomplished by imposing reflecting boundary conditions at infinity, which is a timelike hypersurface that radiation can reach in a finite time. In asymptotically de Sitter spaces one cannot impose such boundary conditions, and we are forced to instead place restrictions on the timescales over which evaporation/thermalization processes occur in order to have a notion of at least quasi-static equilibrium. This approach is of course limited because small black holes shed their mass quickly. 
 
 One solution is to consider an ensemble where the temperature is specified at a finite boundary
 \cite{braden_charged_1990}.   This `isothermal cavity' allows the black hole to come to a stable thermodynamic equilibrium, and has been used to understand the critical behaviour of charged de Sitter black holes both in the standard phase space \cite{carlip_phase_2003}, as well as the extended phase space \cite{simovic_critical_2018-1} (where the cosmological constant acts as a thermodynamic pressure \cite{kubiznak_thermodynamics_2016}).  In the former case   a Hawking-Page-like phase transition in both the asymptotically flat and de Sitter cases is observed \cite{carlip_phase_2003}.
In the latter case, Van der Waals-type phase transitions between small and large black holes were  exhibited, somewhat analogous to their AdS counterparts  \cite{kubiznak_p-v_2012}, but with a new structure emerging in phase space called a `swallowtube' \cite{simovic_critical_2018-1}: a compact region in the extended
phase space outside of which no phase transitions are possible.

Our purpose here is to extend this study to non-linear Born-Infeld electrodynamics.  In anti-de Sitter spacetime, this non-linearity introduces interesting new features beyond the Einstein-Maxwell case
\cite{gunasekaran_extended_2012}: the Gibbs free energy becomes discontinuous in a certain temperature range, indicating a new kind of phase transition between small and intermediate sized black holes. 
In this paper, we demonstrate  that a similar type of phase transition occurs in de Sitter space with a cavity, as well as a reentrant small$\rightarrow$large$\rightarrow$small black hole phase transition. We find that the character of these phase transitions depends critically on the value of the maximal electric field strength in the Born-Infeld theory. We also examine the free energy in an ensemble where the potential is fixed, and find not only metastable reentrant phase transitions (where the radiation phase globally minimizes the free energy), but also a reentrant radiation$\rightarrow$black hole$\rightarrow$radiation transition, the first time such a phase transition has been observed to our knowledge.

Our paper is organized as follows: In Section 2 we briefly review black hole thermodynamics in the extended phase space, and discuss how the isothermal cavity is accounted for. In Section 3 we examine the thermodynamic properties of Born-Infeld-de Sitter black holes, including a discussion of the first law, the phase structure that arises, and an analysis of the vacuum polarization and metric function. Finally, in Section 4 we examine the free energy of charged de Sitter black holes both in the Maxwell and Born-Infeld theories in an ensemble where the potential is fixed, and discuss the phase behaviour present there.

\section{The Action and Thermodynamic Quantities}

Black hole phase transitions have provided us with important insights into quantum gravity in the context of
the AdS/CFT correspondence \cite{maldacena_large_1999}. AdS/CFT relates a $(d+1)$-dimensional gravitational theory in an asymptotically  anti-de Sitter space to a $d$-dimensional conformal field theory on the boundary of the spacetime. Through AdS/CFT, a wide variety of gravitational phenomena can be understood in terms of their CFT counterparts\footnote{For example, a Hawking-Page phase transition in the bulk is dual to a deconfinement transition in the boundary field theory.}, providing a powerful tool for investigating strongly coupled systems such as  quark-gluon plasmas, condensed matter systems, and superfluids  \cite{policastro_ads/cft_2002}.

This has motivated further study into the thermodynamic phase structure of AdS black holes, where extensive effort has been devoted to understanding the {\it extended phase space}, in which the cosmological constant acts as a thermodynamic pressure  \cite{kubiznak_thermodynamics_2016,kastor_enthalpy_2009} via the identification
\begin{equation}\label{press}
P=-\dfrac{\Lambda}{8\pi}
\end{equation}
which can be achieved, for example, with a gauge 3-form potential \cite{creighton_quasilocal_1995}.
The variation of pressure  further requires the presence of a  conjugate volume term in the first law, which leads to a variety of novel thermodynamic behaviour \cite{kubiznak_thermodynamics_2016}, including triple points \cite{altamirano_kerr-ads_2014}, reentrant phase transitions \cite{altamirano_reentrant_2013}, and the emergence of polymer-like \cite{dolan_compressibility_2014,frassino_multiple_2014}
and superfluid-like phase structures \cite{hennigar_superfluid_2017,hennigar_thermodynamics_2017,dykaar_hairy_2017}, in addition to the now commonly observed Van der Waals transition \cite{kubiznak_p-v_2012}. In de Sitter space, the cosmological constant is positive, so the variable $P$ is better thought of as a {\it tension} rather than a pressure.  This introduces qualitatively new behaviour in the phase structure of a black hole, whose implications are the subject of recent exploration  \cite{kubiznak_thermodynamics_2016,simovic_critical_2018-1,mbarek_reverse_2019}.

For the purposes of this paper, the various thermodynamic quantities of interest are derived from the Euclidean action for the metric $g_{\mu\nu}$ over a region $\mathcal{M}$ with boundary $\partial \mathcal{M}$. We are interested in   Einstein-Born-Infeld theory in asymptotically de Sitter space, described by the following bulk action and boundary term:
\begin{equation}
I=-\frac{1}{16\pi}\int_{\mathcal{M}}\!\!d^4x\sqrt{g}\,\big(R-2\Lambda+\mathcal{L}_{\text{BI}}\big)+\frac{1}{8\pi}\int_{\partial\mathcal{M}}\!\!d^3x\sqrt{k}\,\big(K-K_0\big)
\end{equation}
$R$ is the Ricci scalar, $\Lambda$ is the cosmological constant (positive for de Sitter space), $K$ is the trace of the extrinsic curvature of the boundary, and $\mathcal{L}_{\text{BI}}$ is the Born-Infeld Lagrangian, given by:
\begin{equation}\label{LBI}
\mathcal{L}_{\text{BI}}=4b^2\!\left(1-\sqrt{1+\frac{F^{ab}F_{ab}}{2b^2}}\,\right)
\end{equation}
Here, $F_{ab}$ is the usual field strength tensor and $b$ is the Born-Infeld parameter, representing the maximal electromagnetic field strength in the theory, with the limit $b\rightarrow\infty$ corresponding to Maxwell electrodynamics. $b$ can also be related to the string tension $\alpha=(2\pi b)^{-1}$ in string theory, where stretched open strings are created dynamically to suppress the electric field to this maximal value \cite{gibbons_aspects_2001}.

We choose the subtraction term $K_0$ such that the action is normalized to $I=0$ when the mass (and therefore charge) of the black hole vanishes. This makes empty de Sitter space the reference point for the energy, in contrast to the work of Carlip and Vaidya \cite{carlip_phase_2003}, where the boundary term is chosen so that the action vanishes for flat spacetime. As in Einstein-Maxwell-de Sitter gravity \cite{simovic_critical_2018-1}, this choice does not change the qualitative behaviour of the critical phenomena, only the numerical values of various quantities and critical points. It is the more natural choice of reference since empty de Sitter space has the same asymptotics and topology near the boundary as Born-Infeld-de Sitter space.

For the Euclidean metric we take a spherically symmetric ansatz,
\begin{equation}
ds^2=f(y)^2d\tau^2+\alpha(y)^2dy^2+r(y)^2d\Omega^2
\end{equation}
where $y\in[0,1]$ is a compact radial coordinate with $y=0$ corresponding to the black hole horizon ($r(0)=r_+$) and $y=1$ corresponding to the cavity wall ($r(1)=r_c$). The boundary at $y=1$ has topology $S^1\!\times\!S^2$ with the $S^2$ having area $4\pi r_c^2$. Heat flux through the cavity wall is fixed so that its temperature $T=\beta^{-1}$ remains constant. The inverse temperature $\beta$ is related to the proper length of the boundary $S^1$ by $\beta=2\pi f(1)$, where the periodicity in imaginary time $\tau$ is $2\pi$. The details of how to arrive at various other thermodynamic quantities from the action have been summarized elsewhere \cite{braden_charged_1990,simovic_critical_2018-1}. In this ensemble, the energy and entropy are given by
\begin{equation}\label{ESeqs}
E=\dfrac{\partial I_r}{\partial \beta},\qquad S=\beta\left(\dfrac{\partial I_r}{\partial\beta}\right)-I_r
\end{equation}
where $I_r$ is the reduced action. This energy is the mean thermal energy of the black hole with respect to empty de Sitter space, which can be related to the ADM mass of the spacetime after accounting for the gravitational and electrostatic binding energy.

With the energy, temperature and entropy in hand, we can construct the {\it Helmholtz free energy}, $F=E-TS$, which is the thermodynamic potential that is minimized when a system reaches equilibrium at constant temperature. Plotting $F(T)$ for fixed pressure reveals whether any phase transitions occur in the system.

\subsection{The First Law and Smarr Relation}

The first law of thermodynamics for black holes in an asymptotically flat spacetime reads \cite{bardeen_four_1973}
\begin{equation}
dM=\frac{\kappa}{8\pi}dA+\Omega dJ+\phi dQ
\end{equation}
where $\kappa$ is the surface gravity of the black hole, $A$ is the horizon area, $J$ is the angular momentum, $\Omega$ is the angular velocity, $\phi$ is the electrostatic potential, and $Q$ is the electric charge. In the semi-classical regime, one can identify the surface gravity with the black hole temperature $T$, and the area with the entropy associated with horizon degrees of freedom. This leads to the form
\begin{equation}
dM=TdS+\Omega dJ+\phi dQ
\end{equation}
from which the extensive study of black hole thermodynamics emerges.

We will work in the extended phase space where the cosmological constant is treated as a thermodynamic pressure, with the identification given in \eqref{press} between (negative) pressure and (positive) 
$\Lambda$.
The quantity conjugate to $P$ is then interpreted as the thermodynamic volume:
\begin{equation}
V=\left(\dfrac{\partial E}{\partial P}\right)_{S,T}
\end{equation}
This leads to an additional term in the first law corresponding to the variation in the pressure/volume. We must also include a work term associated with changes in the area $A_c$ of the isothermal cavity, whose conjugate is a surface tension/pressure $\lambda$. Finally, the mass parameter $M$ being a function of the Born-Infeld parameter $b$ allows us to define a new quantity  \cite{gunasekaran_extended_2012}
\begin{equation}
\mathcal{B}=\left(\frac{\partial M}{\partial{b}}\right)
\end{equation}
known as the `Born-Infeld vacuum polarization'. This quantity has units of an electric polarization and is required for consistency of the first law and associated Smarr relation \cite{gunasekaran_extended_2012}.

In the case of de Sitter black holes in a cavity, we must identify the energy \eqref{ESeqs}  with the internal energy of the thermodynamic system in order for the first law (and Smarr relation) to be satisfied with the usual definitions of temperature, entropy, and pressure. The above considerations lead us to the following form of the first law
\begin{equation}\label{firstlaw}
dE=TdS-\lambda dA_c-VdP+\phi dQ+\mathcal{B}db
\end{equation}
where the $\Omega dJ$ term has been omitted since we are considering non-rotating black holes only. The action determines the temperature $T$, entropy $S$, and energy $E$ of the system, while the thermodynamic volume, surface tension, and vacuum polarization are determined from the first law. Note that the thermodynamic volume will in general be different from the geometric volume $V=\tfrac{4}{3}\pi r_+^3$ of the black hole.

From the first law, one can also derive the Smarr relation, which in four dimensions is 
\be\label{Smarrmod}
E=2(TS-\lambda A_c+PV)+\Phi Q-\mathcal{B}b
\ee
and can be derived from various scaling arguments \cite{kubiznak_thermodynamics_2016,kastor_enthalpy_2009,smarr_mass_1973}. This relation is broadly applicable as it holds for both asymptotically AdS and dS spacetimes, is valid in any dimension, and is also satisfied by more exotic objects like black rings and black branes \cite{kastor_enthalpy_2009}.

\section{Born-Infeld-de Sitter Black Holes}

We now move to the discussion of the Born-Infeld-de Sitter black hole. In Schwarzschild coordinates, the metric function takes the form
\begin{equation}
ds^2=-N(r)^2dt^2+\dfrac{dr^2}{N(r)^2}+r^2d\Omega^2
\end{equation}
where $N(r)$ is determined by the Hamiltonian constraint and is given by:
\begin{equation}
N(r)=1-\dfrac{m}{r}-\dfrac{\Lambda r^2}{3}+\dfrac{2b^2}{r}\int_r^{\infty}\left(\sqrt{r^4+\dfrac{q^2}{b^2}}-r^2\right)dr
\end{equation}
In these coordinates, the gauge field can be written
\begin{equation}
A_t=\frac{q}{r}\ _2F_1\!\left(\dfrac{1}{4},\dfrac{1}{2},\dfrac{5}{4},\dfrac{-q^2}{b^2r^4}\right)
\end{equation}
giving a finite radial electric field
\begin{equation}
E(r)=\frac{q}{\sqrt{r^4+\frac{q^2}{r^2}}}
\end{equation}
where $_2F_1(a,b,c,z)$ is the hypergeometric function.
The `integral' form of the metric function \eqref{N(r)} is valid for all $\Lambda$, $b$, $q$ and $r>0$; it  can also be written in terms of the Gauss hypergeometric function as
\begin{equation}\label{N(r)}
N(r)=1-\dfrac{m}{r}-\dfrac{\Lambda r^2}{3}+\dfrac{2b^2r^2}{3}\left(1-\sqrt{1+\dfrac{q^2}{b^2r^4}}\right)+\dfrac{4q^2}{3r^2} \ _2F_1\!\left(\dfrac{1}{4},\dfrac{1}{2},\dfrac{5}{4},\dfrac{-q^2}{b^2r^4}\right)
\end{equation}
where the series expansion of $_2F_1(a,b,c,z)$ is convergent for $|z|>1$, i.e. $r>\sqrt{q/b}$. The reduced action for this spacetime can then be determined:
\begin{align}
I_r=&\beta r_c\left[\left(1-\frac{\Lambda r_c^2}{3}\right)^{\!1/2}-\left(\frac{(2b^2-\Lambda)(r_c^3-r_+^3)+3(r_c-r_+)-2br_c\sqrt{b^2r_c^4+q^2}}{3r_c}\right.\right.\nonumber\\
&\!\!\!\left.\frac{+2br_+\sqrt{b^2r_+^4+q^2}}{3r_c}-\frac{4q^2\left( _2F_1\!\left(\frac{1}{4},\frac{1}{2},\frac{5}{4},\frac{-q^2}{b^2r_+^4}\right) r_c-\!\ _2F_1\!\left(\frac{1}{4},\frac{1}{2},\frac{5}{4},\frac{-q^2}{b^2r_c^4}\right) r_+\right)}{3r_c^2r_+}\Bigg)^{\!1/2}\,\right]-\pi r_+^2
\end{align}
The inverse temperature (and therefore temperature) is found by extremizing the action with respect to $r_+$ and solving for $\beta$, giving
\begin{equation}\label{temp}
T=\frac{\sqrt{3}\,r_c \left(1+r_+^2 \left(2 b^2-\Lambda\right)-2\,b \sqrt{b^2 r_+^4+q^2}\right)}{4\pi\sqrt{r_+}\,X}
\end{equation}
where we have defined 
\begin{equation}
X\equiv\sqrt{r_c\, r_+\! \left(3 (r_c-r_+)+\left(2 b^2-\Lambda\right) \left(r_c^3-r_+^3\right)-2 b\! \left(r_c \sqrt{b^2 r_c^4+q^2}-r_+ \sqrt{b^2 r_+^4+q^2}\right)\right)-\textsf{F}}
\end{equation}
\begin{equation}
\textsf{F} \equiv 4q^2\left[ _2F_1\!\left(\frac{1}{4},\frac{1}{2},\frac{5}{4},\frac{-q^2}{b^2r_+^4}\right) r_c-\!\ _2F_1\!\left(\frac{1}{4},\frac{1}{2},\frac{5}{4},\frac{-q^2}{b^2r_c^4}\right) r_+\right]
\end{equation}
In the limit $b\rightarrow\infty$, this reduces to the Maxwell case \cite{simovic_critical_2018-1},
\begin{equation}
T=\dfrac{1-\frac{q^2}{r_+r_c}-\frac{\Lambda}{3}\big(r_c^2+r_cr_++r_+^2\big)+\left(1-\frac{r_+}{r_c}\right)\left(\frac{\Lambda}{3}(r_c^2+2r_+r_c)-\frac{q^2}{r_+^2}\right)}{4\pi r_+\sqrt{\left(1-\frac{r_+}{r_c}\right)\Big(1-\frac{q^2}{r_+r_c}-\frac{\Lambda}{3}\big(r_c^2+r_cr_++r_+^2\big)\Big)}}.
\end{equation}
and taking further the limits $\Lambda\rightarrow 0$, $q\rightarrow0$, and $r_c\rightarrow\infty$ gives the familiar result $T=1/4\pi r_+$ for the asymptotically flat Schwarzschild black hole. The entropy is
\begin{equation}\label{entropy}
S=\beta\dfrac{\partial I_r}{\partial\beta}-I_r=\pi r_+^2
\end{equation}
and finally the energy is
\begin{align}
E=\dfrac{\partial I_r}{\partial\beta}=&\ r_c\left[\left(1-\frac{\Lambda r_c^2}{3}\right)^{\!1/2}-\left(\frac{2b(r_c^3-r_+^3)^2-\Lambda(r_c^3-r_+^3)+3(r_c-r_+)-2br_c\sqrt{b^2r_c^4+q^2}}{3r_c}\right.\right.\nonumber\\
&\!\left.\frac{+2br_+\sqrt{b^2r_+^4+q^2}}{3r_c}-\frac{4q^2\left( _2F_1\!\left(\frac{1}{4},\frac{1}{2},\frac{5}{4},\frac{-q^2}{b^2r_+^4}\right) r_c-\!\ _2F_1\!\left(\frac{1}{4},\frac{1}{2},\frac{5}{4},\frac{-q^2}{b^2r_c^4}\right) r_+\right)}{3r_c^2r_+}\Bigg)^{\!1/2}\,\right]\nonumber\\
=&\frac{1}{3} \left(r_c \sqrt{9-3 \Lambda r_c^2}-X\sqrt{3/r_+} \right)
\label{intE}
\end{align}
One can check that in the limit $b\rightarrow\infty$ these quantities reduce to those obtained in \cite{simovic_critical_2018-1}.

\subsection{The First Law}

With the energy $E$, temperature $T$, and entropy $S$ as defined above, we can determine the thermodynamic volume $V$, surface tension $\lambda$, and vacuum polarization $\mathcal{B}$ from the first law. In order for \eqref{firstlaw} to hold, the surface tension must be
\begin{align}
\lambda=\ &\frac{1}{8\pi r_c}\left(\frac{3-2 \Lambda r_c^2}{\sqrt{9-3 \Lambda r_c^2}}+\frac{\textsf{F}-r_c {r_+} \left(\left(2 b^2-\Lambda\right) \left(4 r_c^3-{r^3_+}\right)+3 (2 r_c - {r_+})\right)}{2 \sqrt{3} r_c  {\sqrt{r_+}} X}\right.\nonumber\\
&\qquad\qquad  \left.-\frac{2 b \left(4 r_c \sqrt{b^2 r_c^4+q^2}- {r_+} \sqrt{b^2  {r^4_+}+q^2}\right)}{2 \sqrt{3} r_c \sqrt{ {r_+}} X}\right)\ ,
\end{align}
the thermodynamic volume is
\begin{align}\label{vol}
V=\frac{4 \pi  r_c \left(r_c^3 \sqrt{r_+}-\frac{r_c^2 X}{\sqrt{3-\Lambda r_c^2}}-r_+^{7/2}\right)}{\sqrt{3}\, X}\ ,
\end{align}
the electric potential is
\begin{equation}\label{ePot}
\phi=\frac{\sqrt{3}\, \textsf{F}}{4 q \sqrt{r_+} X}\ ,
\end{equation}
and the vacuum polarization is
\begin{equation}
\mathcal{B}=\frac{2 r_c r_+ \left[q^2 \left(\frac{r_c}{\sqrt{b^2 r_c^4+q^2}}-\frac{r_+}{\sqrt{b^2 r_+^4+q^2}}\right)+b^2 \left(\frac{r_c^5}{\sqrt{b^2 r_c^4+q^2}}-\frac{r_+^5}{\sqrt{b^2 r_+^4+q^2}}\right)-b \left(r_c^3-r_+^3\right)\right]+\dfrac{\textsf{F}}{4 b}}{\sqrt{3r_+}\, X}\ .
\end{equation}
With these definitions it is straightforward to verify that  
\begin{equation}
dE=TdS-\lambda dA_c-VdP+\phi dQ+\mathcal{B}db
\end{equation}
where the sign of the $VdP$ term is negative in order to ensure that the thermodynamic volume is positive in the appropriate region. 

\begin{figure}[h]
	\includegraphics[width=0.49\textwidth]{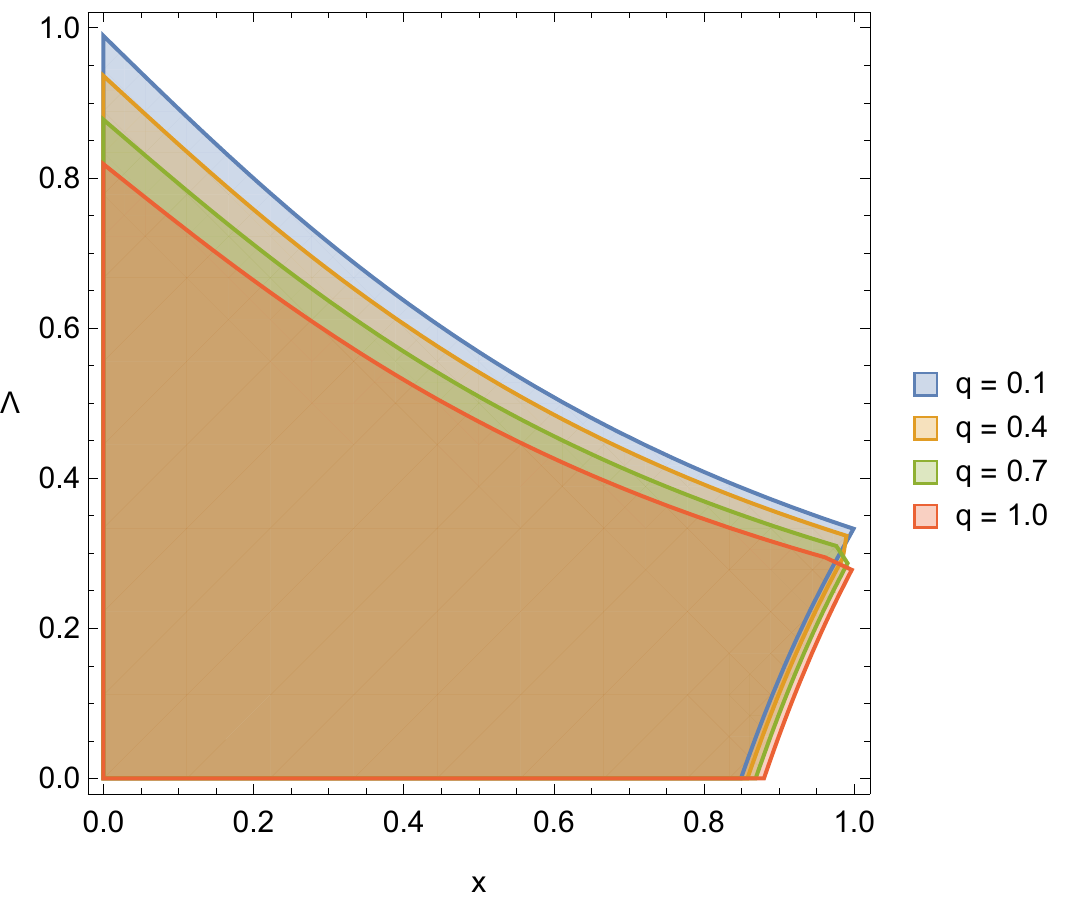}\quad\includegraphics[width=0.49\textwidth]{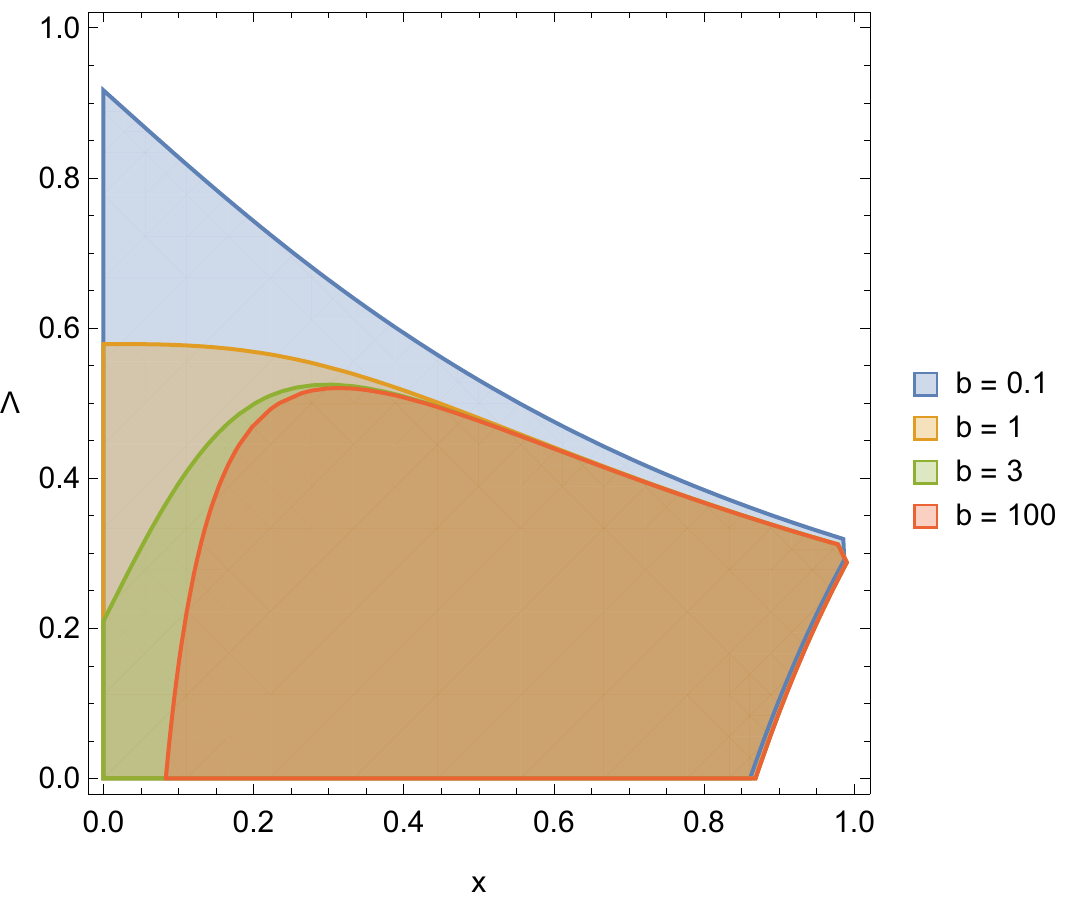}
	\caption{Regions of positivity for the thermodynamic volume $V$ as a function of $x$ and $\Lambda$ with fixed cavity radius $r_c=\sqrt{3}$. Shaded regions indicate positivity. \textbf{Left:} Varying charge with $b=0.1$. \textbf{Right:} Varying $b$ with $q=0.5$.}
\end{figure}

\begin{figure}[h]
	\includegraphics[width=0.49\textwidth]{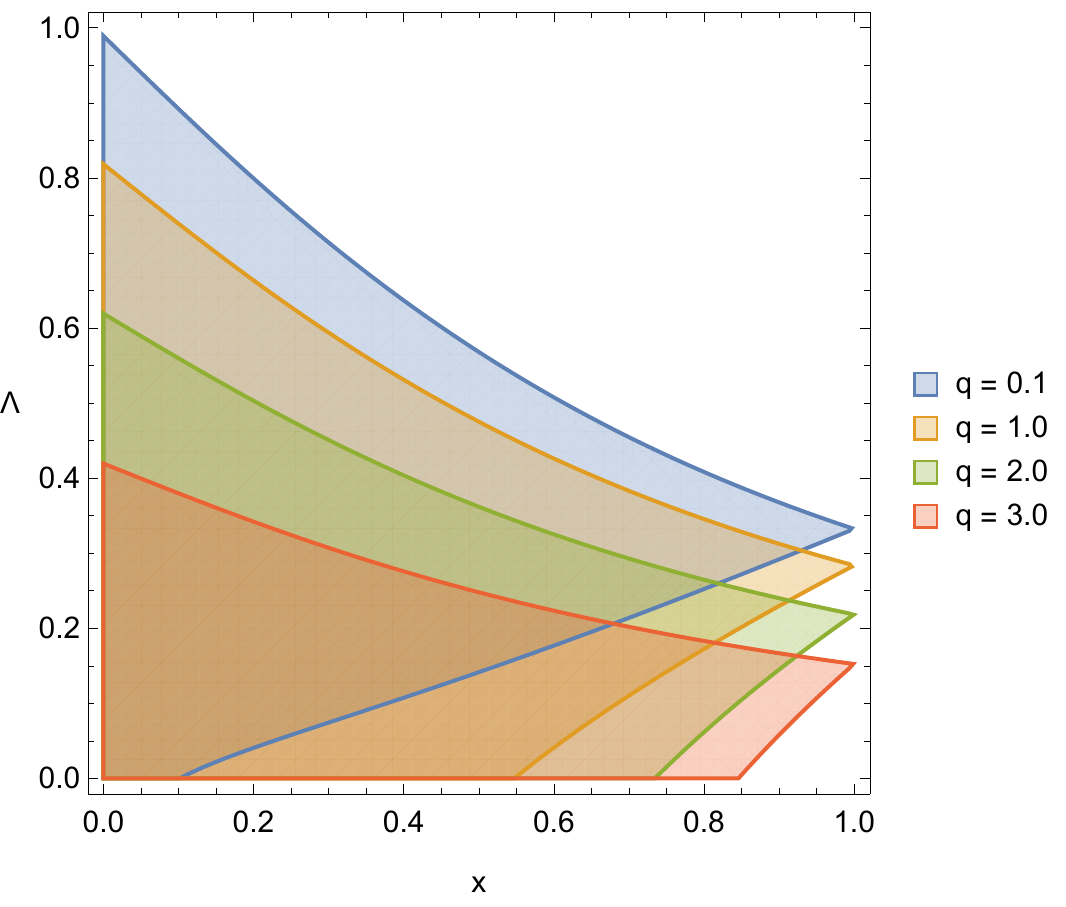}\quad\includegraphics[width=0.49\textwidth]{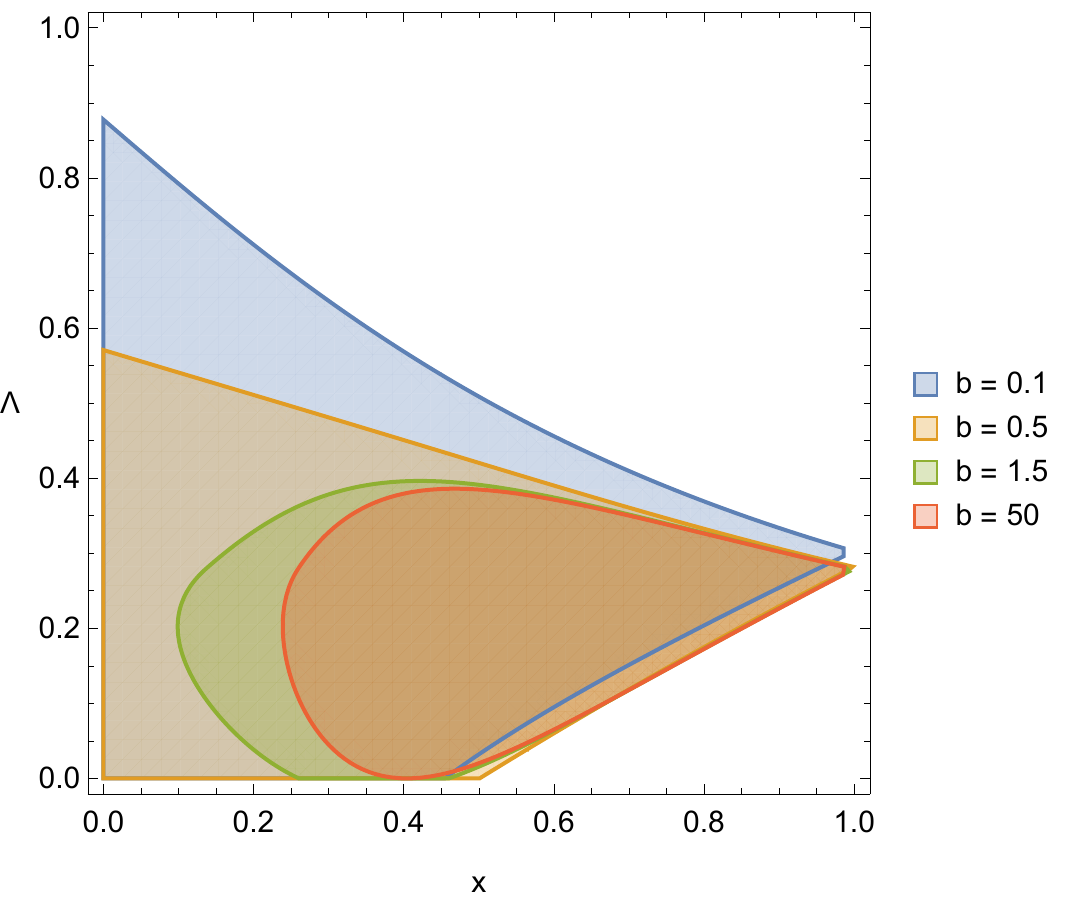}
	\caption{Regions of positivity for the surface tension $\lambda$ as a function of $x$ and $\Lambda$ with fixed cavity radius $r_c=\sqrt{3}$. Shaded regions indicate positivity, where $\lambda$ should be regarded as a surface pressure. \textbf{Left:} Varying charge with $b=0.1$. \textbf{Right:} Varying $b$ with $q=0.7$. }
\end{figure}

Figure 1 shows regions in parameter space where $V$ is positive, in terms of the dimensionless ratio $x\equiv r_+/r_c\in[0,1]$. These regions correspond to choices of parameters \{$m$, $\Lambda$, $q$, $b$\} which lead to a sensible notion of volume. We have implicitly restricted our analysis to regions where both $r_+<r_c$ and $r_c<r_{\text{cosmo}}$\footnote{The cosmological horizon, $r_{\text{cosmo}}$, is located at the largest real root of $N(r)=0$.}, the second restriction being necessary to avoid acausal influence of the isothermal cavity on the black hole. Outside the shaded regions one cannot interpret \eqref{vol} as a volume in the traditional thermodynamic sense. One can see on the right side of Figure 1 that as $b$ becomes large, the volume behaves as it does in the Maxwell case, as expected. Figure 2 depicts regions where $\lambda$, which appears in the first law as a work term related to changes in the cavity size, is positive. Outside of the shaded regions (some of which overlap with $V>0$ regions) $\lambda$ is negative, and thus better thought of as a surface {\it tension}.
Within the shaded regions  $\lambda$ is positive, and so should be regarded as a 
surface pressure.

\subsection{Vacuum Polarization and Metric}

Born-Infeld theory has the feature that two distinct types of black hole solutions exist, depending on the values of $b$ and $q$. This is demonstrated in Figure 3, where we plot the metric function \eqref{N(r)} for fixed mass and varying $b$, as well as fixed $b$ and varying mass. There is a {\it marginal} case (indicated in green) separating the two types of solutions, which occurs when $N(r)$ attains a finite value at the origin. The condition for this to occur is
\begin{equation}\label{mm}
m_m=\dfrac{\sqrt{b\,q^3}\,\Gamma\!\left(\frac{1}{4}\right)^2}{3\sqrt{\pi}}\; .
\end{equation}
Spacetimes with $m>m_m$ are Schwarzschild-de Sitter-like, having a singularity at $r=0$, a black hole horizon at $r=r_+$, and a cosmological horizon at $r=r_{\text{cosmo}}$ (beyond the cavity radius $r_c$). Spacetimes with $m<m_m$ are Reissner-Nordstrom-de Sitter-like (and approach this solution as $b\to \infty$), with up to three distinct horizons: the inner horizon at $r=r_-$, the outer black hole horizon at $r=r_+$, and the cosmological horizon at $r=r_{\text{cosmo}}$. The small-$r$ behaviour is similar to that of the Born-Infeld-AdS case examined in \cite{gunasekaran_extended_2012}. However the positive cosmological constant changes the large-$r$ behaviour; even though the cosmological horizon is `hidden' behind the cavity, spacetime is still de Sitter-like up to the cavity radius.
\begin{figure}[h]
	\includegraphics[width=0.485\textwidth]{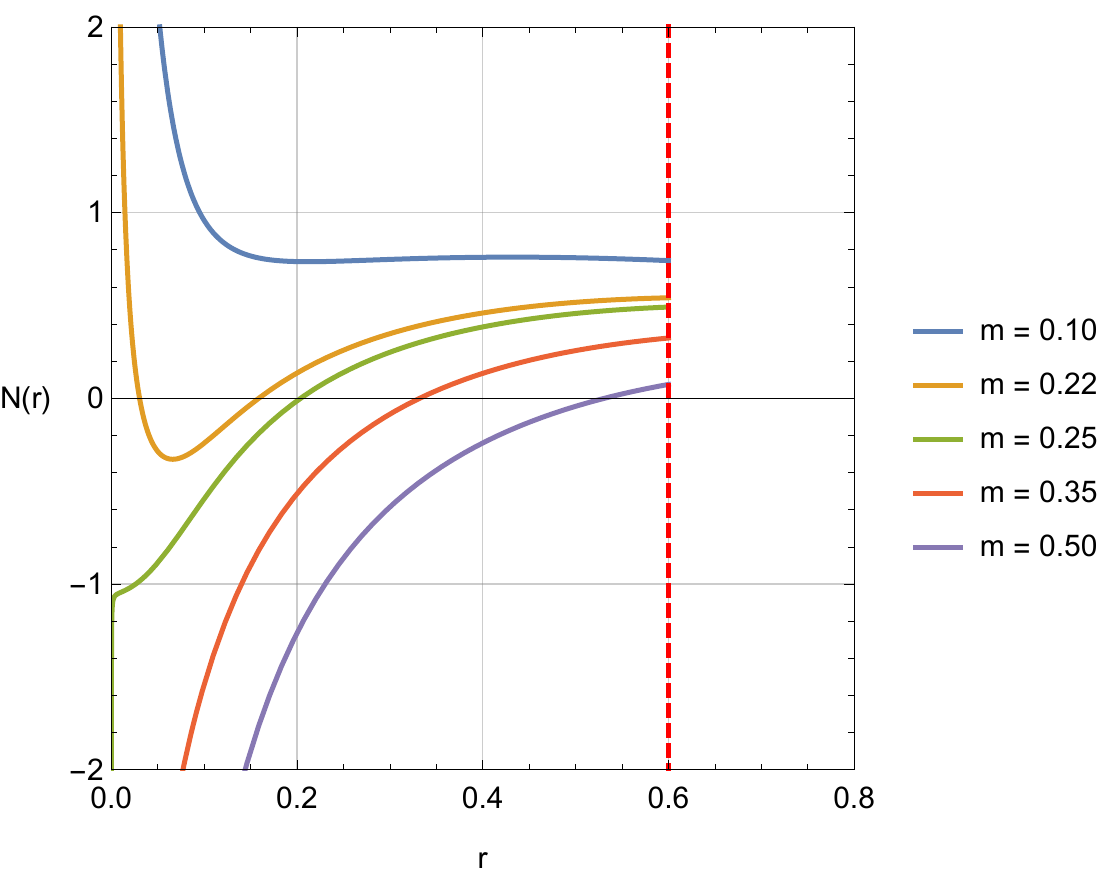}\quad\includegraphics[width=0.49\textwidth]{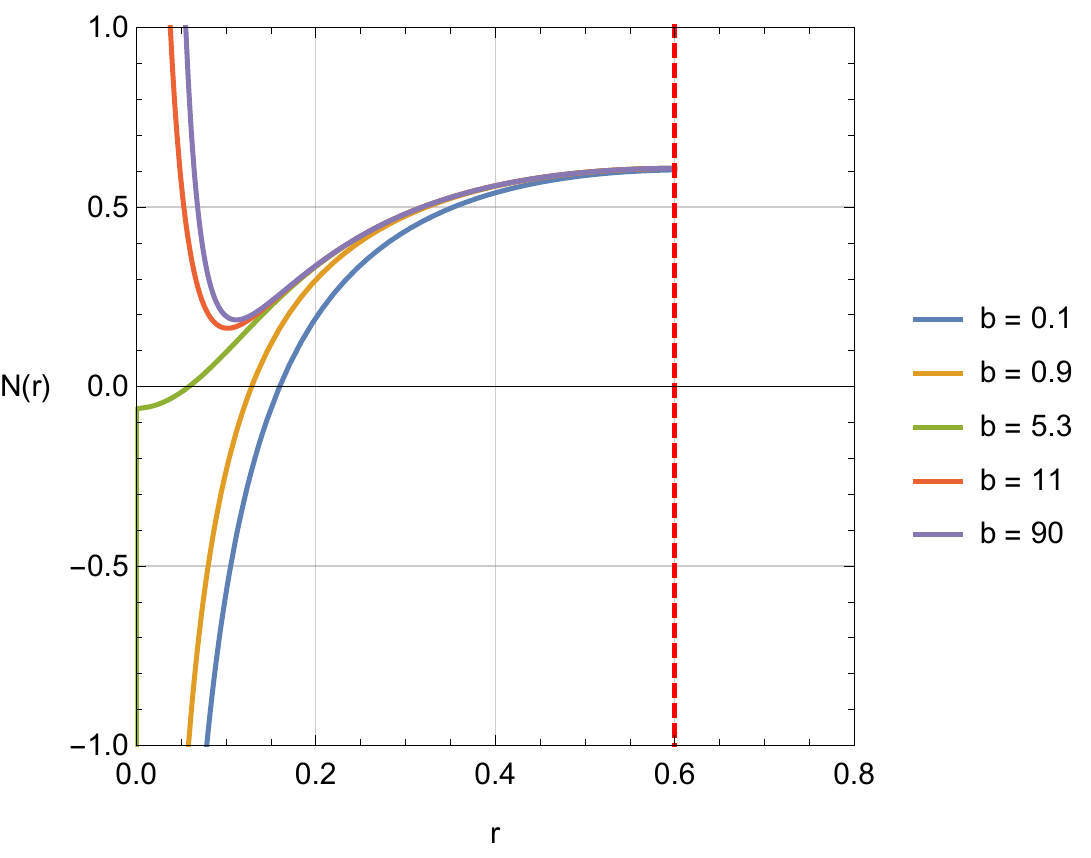}
	\caption{The metric function $N(r)$ as a function of $r$ for $q=0.1$, $\Lambda=1$, and $r_c=0.6$; the cavity radius is 
	indicated by the red dashed line. \textbf{Left:} Fixed vacuum polarization ($b=10.22$) and varying mass parameter $m$. \textbf{Right:} Fixed mass ($m=0.18$) and varying $b$. In both figures the marginal case appears in green.
	}
\end{figure}

\begin{figure}[h]
	\includegraphics[width=0.495\textwidth]{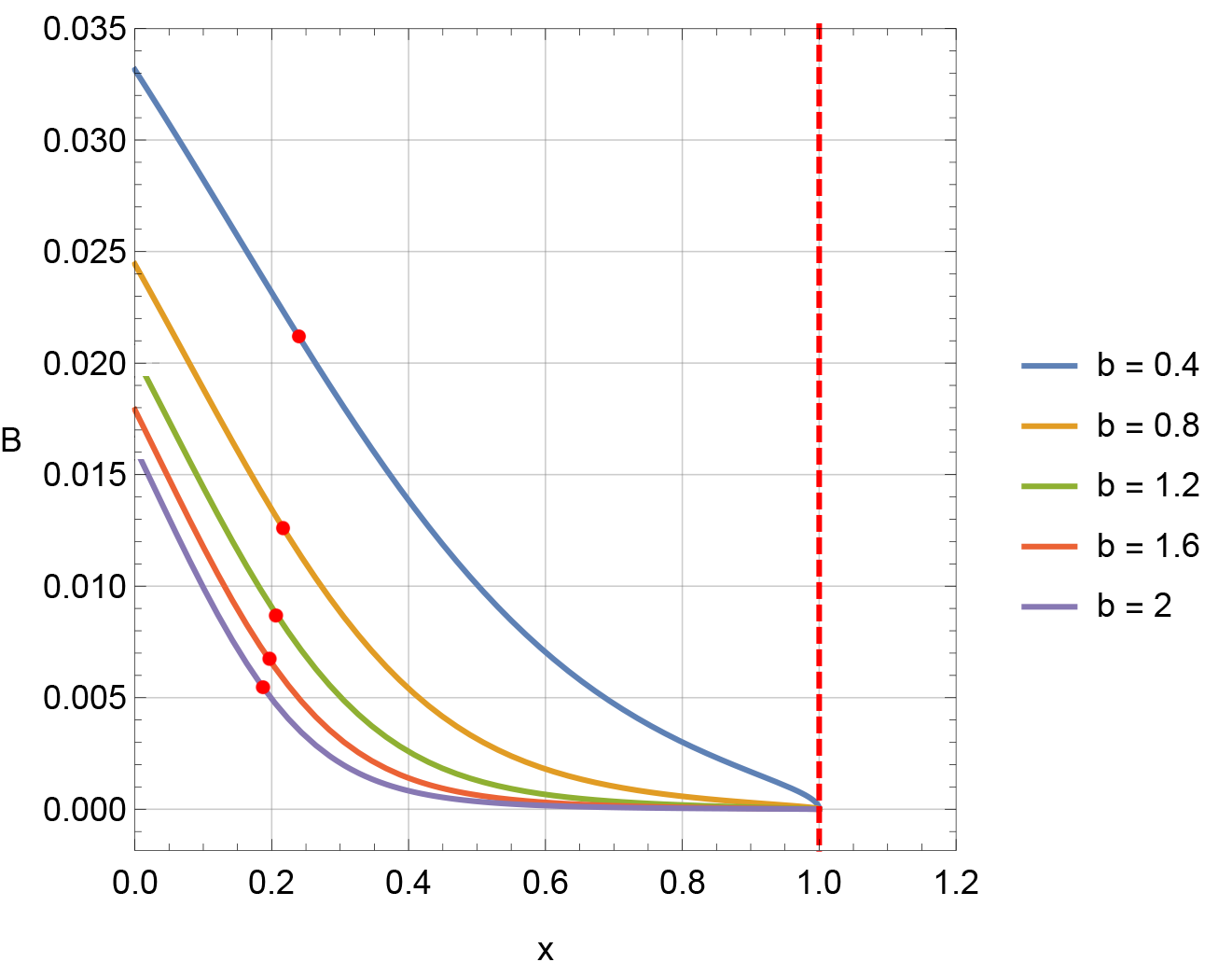}\quad\includegraphics[width=0.475\textwidth]{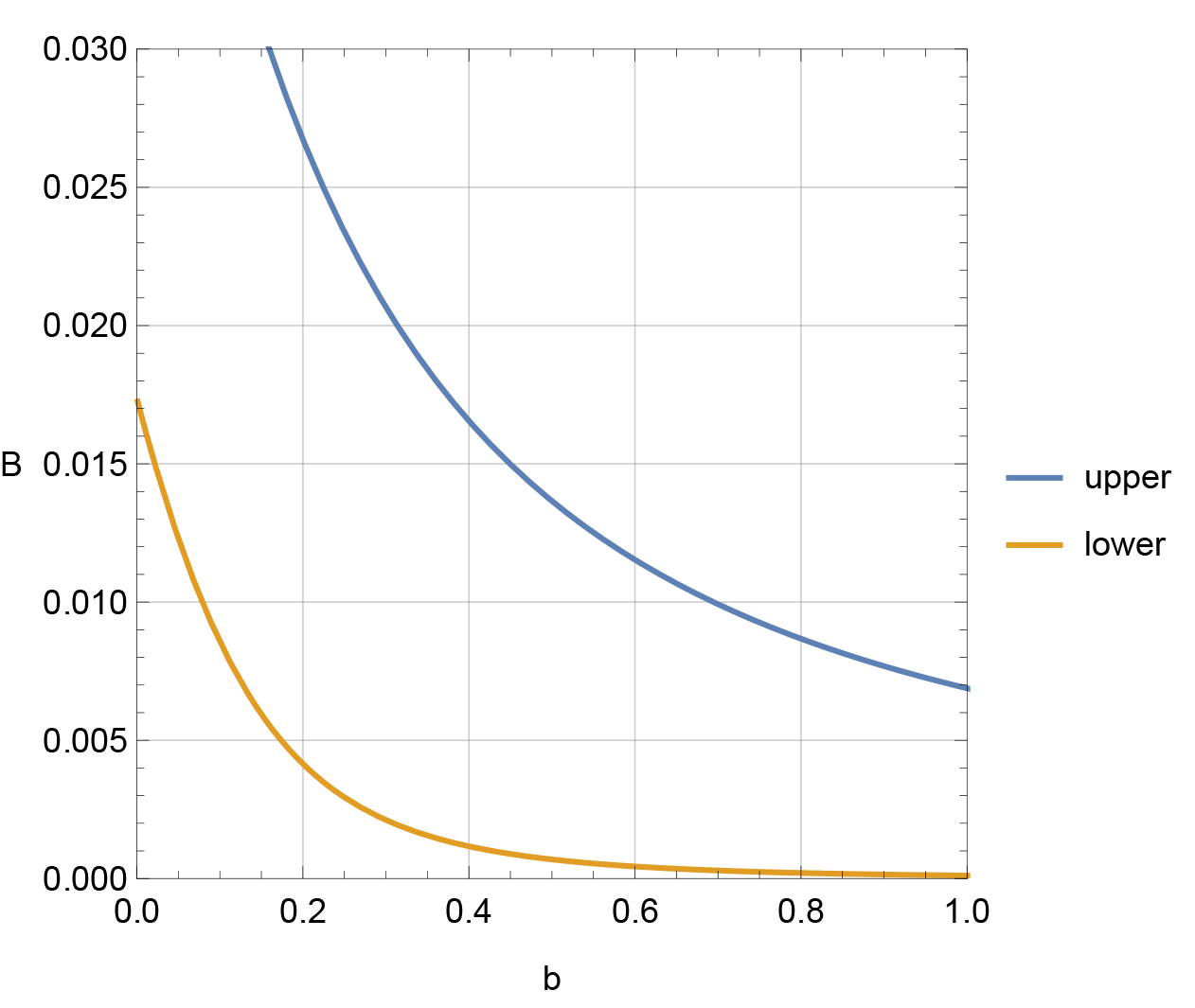}
	\caption{Behaviour of the vacuum polarization for $q=0.1$, $\Lambda=1$, and $r_c=0.6$; the cavity radius is 	indicated by the red dashed line. \textbf{Left:} $\mathcal{B}$ as a function of $x\equiv r_+/r_c$ for various $b$. The red dots indicate the marginal mass along each line. \textbf{Right:} $\mathcal{B}$ as a function of $b$ for fixed $T=0.5$ ($\mathcal{B}-b$ isotherms), with two branches.
}
\end{figure}

The conjugate $\mathcal{B}$ to the maximal electric field strength $b$ in the first law has units of polarization per unit volume, and is referred to as the {\it Born-Infeld vacuum polarization}. Its behaviour is shown in Figure 4. On the left, we see that for any $r<r_c$, the vacuum polarization approaches zero as $b$ increases (corresponding to the Maxwell limit), and appears to reach a finite value at the origin. However we cannot draw conclusions about the small-$r$ behaviour since \eqref{LBI} restricts us to working in regions where $r>\sqrt{q/b}$. The red dots on each line mark where the marginal mass \eqref{mm} is achieved; to the left of this dot the solution is Reissner-Nordstrom-de Sitter-like and to the right it is Schwarzschild-de Sitter-like.
 
Also of note is the fact that $\mathcal{B}$ vanishes at $r_c$ for any value of $b$, implying that the cavity itself does not support a vacuum polarization. Looking at the right diagram in Figure 4, we see that there are two branches of $\mathcal{B}-b$ isotherms. The upper branch corresponds to a negative thermodynamic volume $V$, while the lower branch has $V>0$, so we take the lower branch to be the `physical' one. In both branches the vacuum polarization approaches zero as we approach the Maxwell limit $b\rightarrow\infty$.

\subsection{Helmholtz Free Energy and Phase Transitions}

We now turn to the phase structure of the Born-Infeld-de Sitter black hole. The quantity of interest is the Helmholtz free energy, $F=E-TS$, which is minimized by the equilibrium state of the system. Plotting $F$ as a function of $T$ for fixed $P$ (or $\Lambda$) will reveal the presence of any phase transitions in the system, generally indicated by discontinuities in the free energy. Using \eqref{temp}, \eqref{entropy}, and \eqref{intE}
we have
\begin{align}\label{F}
F(r_+,\Lambda)&=r_c \sqrt{1-\frac{\Lambda r_c^2}{3}}+\frac{\sqrt{3} r_c r_+^{3/2} \left(1+ \left(2 b^2-\Lambda\right)r_+^2-2 b \sqrt{q^2+b^2 r_+^4}\right)}{4 X}-\frac{X}{\sqrt{3 r_+}}
\end{align}
where $F$ is understood to also depend on $q$, $b$, and $r_c$, which are fixed. We plot $F(T)$ parametrically using $r_+$ as the parameter. This is shown in Figures 5 and 6.
\begin{figure}[h]
	\includegraphics[width=0.47\textwidth]{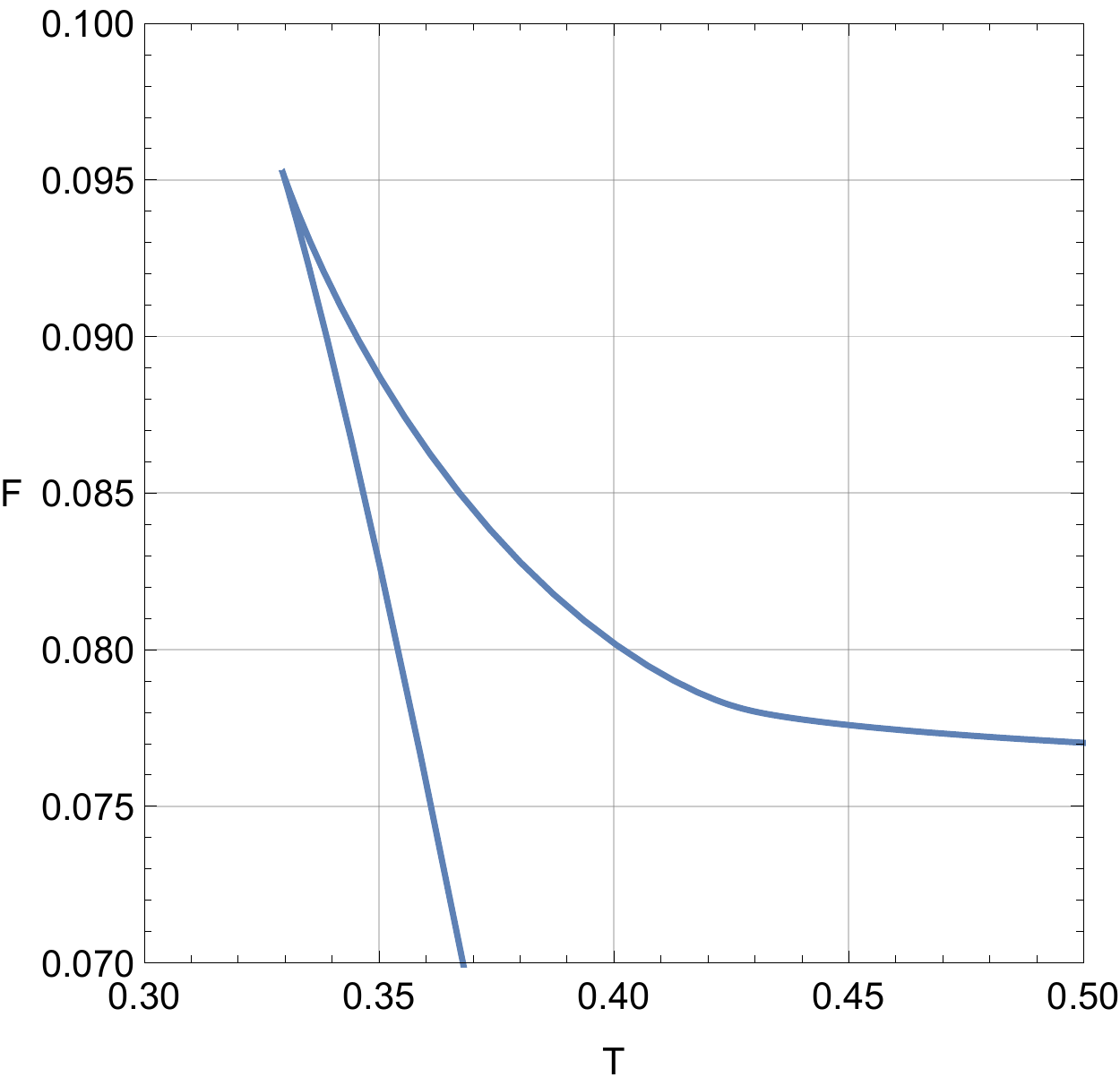}\qquad\includegraphics[width=0.46\textwidth]{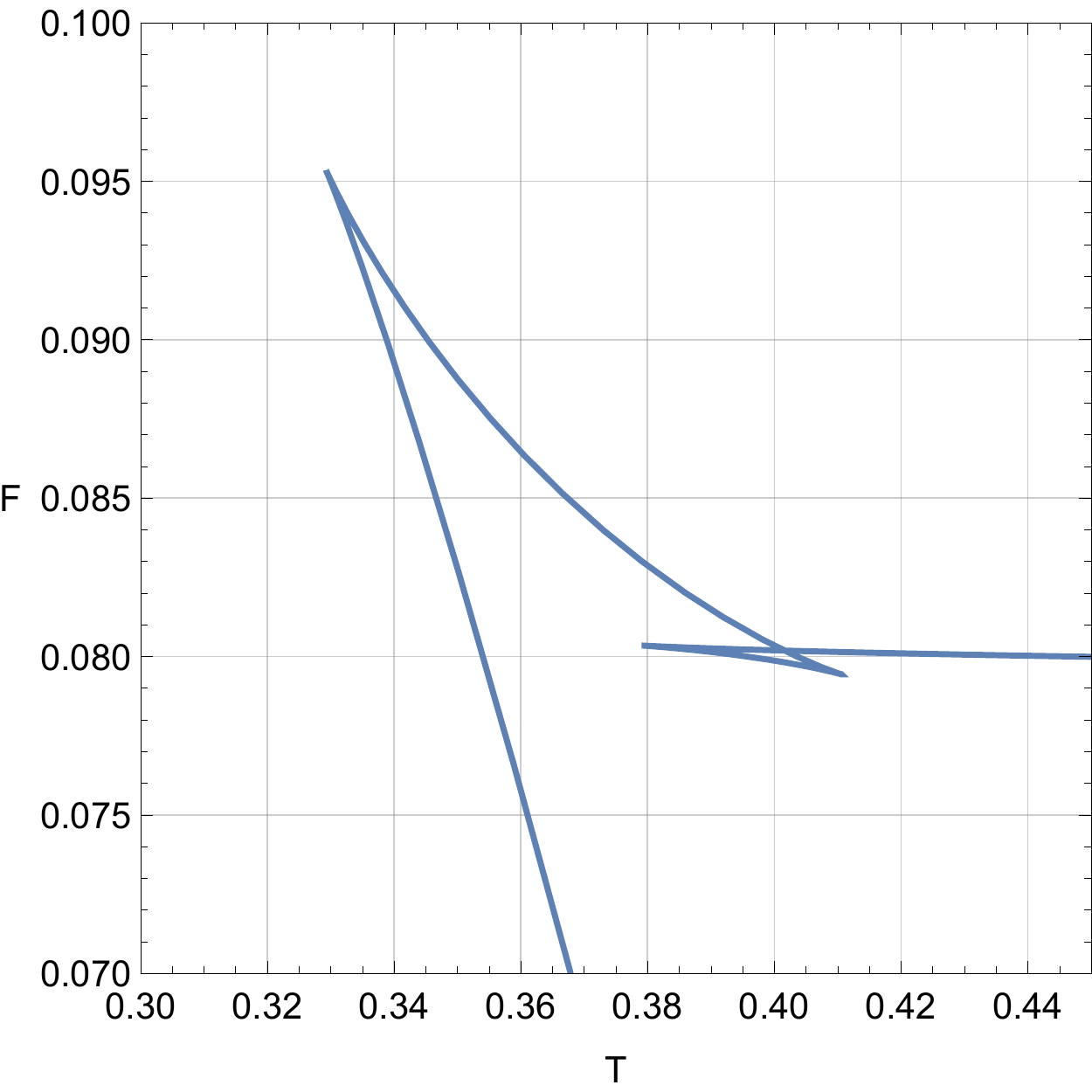}
	\caption{Helmholtz free energy of the Born-Infeld-de Sitter black hole for fixed cavity size ($r_c=0.6$), pressure ($P=-0.0025$), and charge ($q=0.1$). \textbf{Left:} $b=4$. \textbf{Right:} $b=4.42$.}
\end{figure}

 There are a total of 4 distinct critical values of $b$ marking points where qualitative differences in the phase behaviour occur. We refer to these as \{$b_{c1},b_{c2},b_{c3},b_{c4}$\} and note that the exact values depend on $q$, $r_c$, and $\Lambda$ and must be found numerically. On the left of Figure 5, we see that for `small' values of the maximal electric field strength ($b<b_{c1}$), the free energy resembles that of an uncharged AdS black hole, where a Hawking-Page phase transition normally occurs. In this case however, we have chosen an ensemble where the charge is fixed, so there can be no transition from a black hole to empty space, and only a large black hole exists above the minimum temperature indicated by the red line.  To see a possible phase transition, we must plot the free energy at fixed potential, $G=M-TS-\Phi q$, which we do in Section 4. Above the critical value $b_{c1}$ (in this case $b_{c1}\approx 4.38$) a kink forms in the free energy, which traces out an inverted swallowtail in $F-T-P$ space, similar to the one observed in \cite{gunasekaran_extended_2012}. This is shown in the right-hand figure in Figure 5. However, for values of $b$ in the range ($b_{c1},b_{c2}$) this kink does not intersect the lower large black hole branch, which still minimizes the free energy, and therefore does not indicate the presence of a phase transition.
 
 \begin{figure}[h]
 	\includegraphics[width=0.47\textwidth]{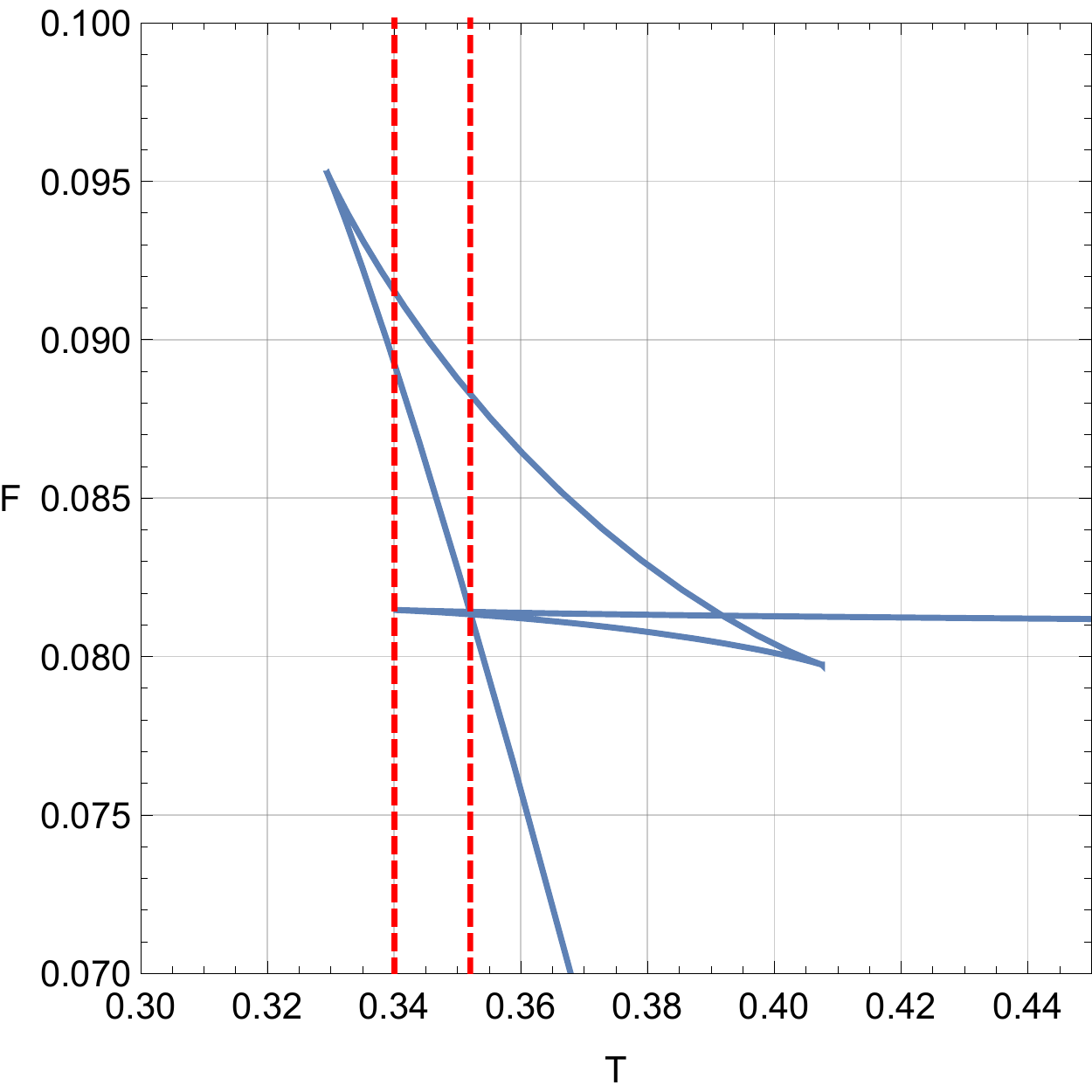}\qquad\includegraphics[width=0.47\textwidth]{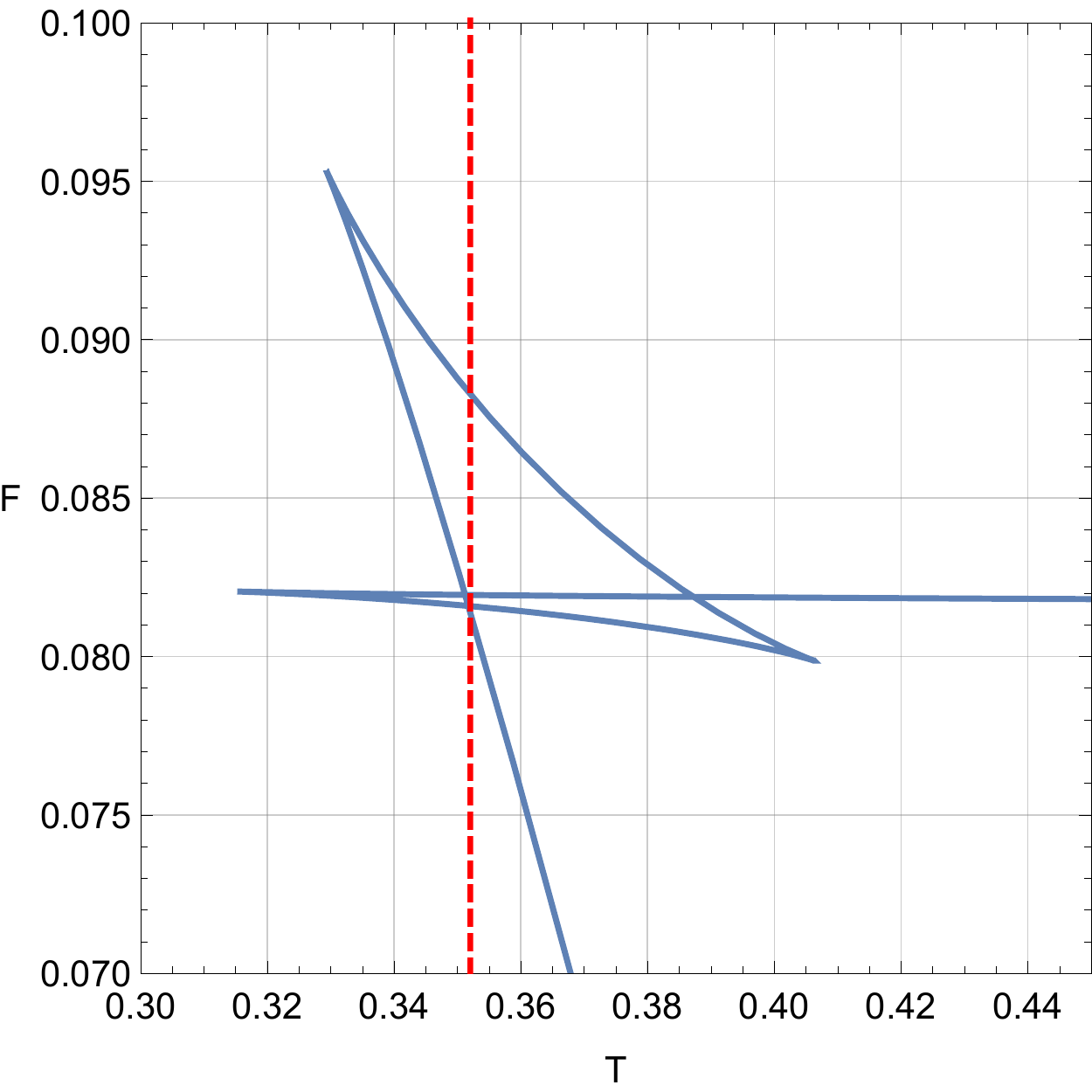}
 	\caption{Helmholtz free energy of the Born-Infeld-de Sitter black hole for fixed cavity size ($r_c=0.6$), pressure ($P=-0.0025$), and charge ($q=0.1$). On the left we have $b=4.57565$ and on the right $b=4.65$. The red dashed lines mark the temperature at which phase transitions occur.} 
 \end{figure}
 
 In Figure 6, we see that when $b$ exceeds the second critical value ($b>b_{c2}$), a `star' pattern characteristic of a reentrant phase transition forms. Examining the figure on the left, as the temperature increases the first dashed line corresponds to a zeroth-order large$\rightarrow$small black hole phase transition, whereas the second dashed line corresponds to the usual first-order small$\rightarrow$large transition.  When $b>b_{c3}$ (as in the right side of Figure 6), the small black hole branch always has lower free energy than the upper large black hole branch, so the first phase transition is washed out and only the small$\rightarrow$large black hole transition at the second dashed line remains. Finally, when $b$ is above the fourth critical value $b>b_{c4}$, the phase structure of a charged Schwarzschild-de Sitter black hole emerges, as expected since the limit $b\rightarrow\infty$ corresponds to the usual Einstein-Maxwell action. Indeed in this limit we recover all of the phenomena and qualitative behaviour seen in previous work \cite{simovic_critical_2018-1}, with only small quantitative changes to the various critical values when $b$ is very large but finite. To summarize, there are five regions separated by four critical values of $b$:
\\

{\small
\begin{tabular}{rl}
	\setlength{\itemsep}{-8pt}
	$0<b<b_{c1}$\ : & No phase transitions. Large black hole phase globally minimizes $F$. \\
	$b_{c1}<b<b_{c2}$\ : & No phase transitions. Unstable inverted swallowtail region forms.\\
	$b_{c2}<b<b_{c3}$\ : & Reentrant phase transition from large$\rightarrow$small$\rightarrow$large black hole.\\
	$b_{c3}<b<b_{c4}$\ : & Phase transition from small$\rightarrow$large black hole, with a minimum temperature.\\
	$b_{c4}<b<\infty\,$\ : & Phase transition from small$\rightarrow$large black hole, with no minimum temperature.\\
\end{tabular}
}
\vspace{12pt}

In the Maxwell case examined in \cite{simovic_critical_2018-1}, we demonstrated the presence of a unique `swallowtube' structure in phase space which results from the presence of the isothermal cavity.   This tube represented a compact region in phase space where a small/large black hole phase transition occurs, the pressure being bounded by an upper and lower critical value $P\in\{P_{\text{min}},P_{\text{max}}\}$ outside of which the phase transition disappears. A similar, albeit more complicated structure appears when Born-Infeld electrodynamics is introduced. For large values of $b$ we recover the swallowtube seen in \cite{simovic_critical_2018-1}, with a slightly modified shape (the identical shape being achieved in the limit $b\rightarrow\infty$). However once $b$ falls below a critical value, the shape of this tube qualitatively changes, such that each constant-pressure slice resembles one of the two cases shown in
Figure 6. Depending on the exact choice of parameters, these slices may contain a reentrant phase transition (as in Figure 6a), or a regular small$\rightarrow$large transition (as in Figure 6b), and always have the property that there are no black holes with $T\approx 0$, unlike in the Maxwell case where black holes exist down to $T=0$. This difference is shown in Figure 7, with the tube structure being represented as a series of constant pressure slices in $F-T-P$ space. Furthermore, for any choice of $q$, $r_c$, and $b$ where a reentrant phase transition is present in one of the slices, the tube terminates at $P=0$. If there is no reentrant phase transition, one can always choose parameters such that the tube pinches off at a non-zero pressure as it does in the Maxwell case. Note finally that if $b$ is small enough, then for no choice of $q$, $r_c$, and $\Lambda$ is the swallowtube present.

\begin{figure}[h]
\centering
\includegraphics[width=0.5\textwidth]{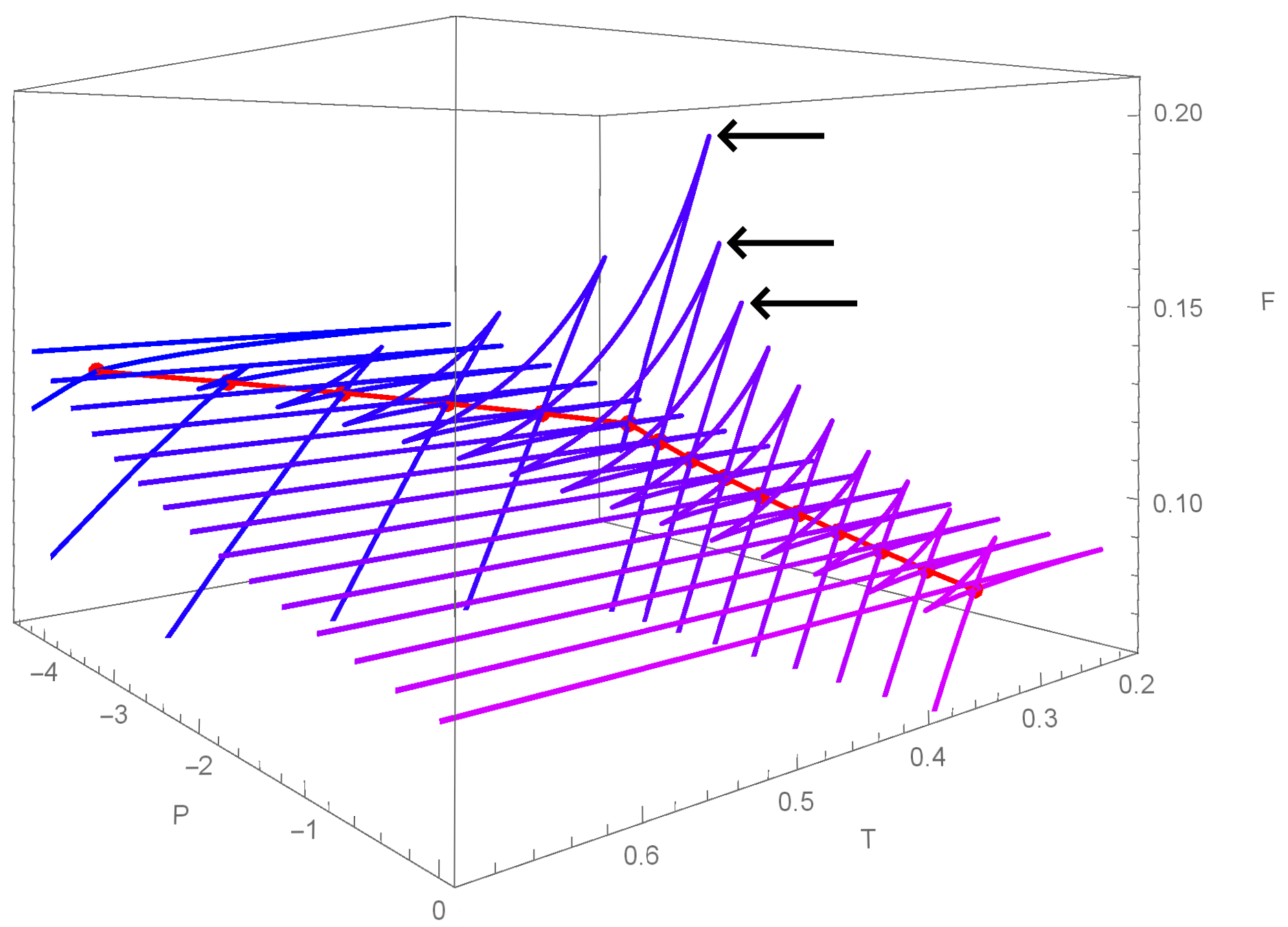}\includegraphics[width=0.5\textwidth]{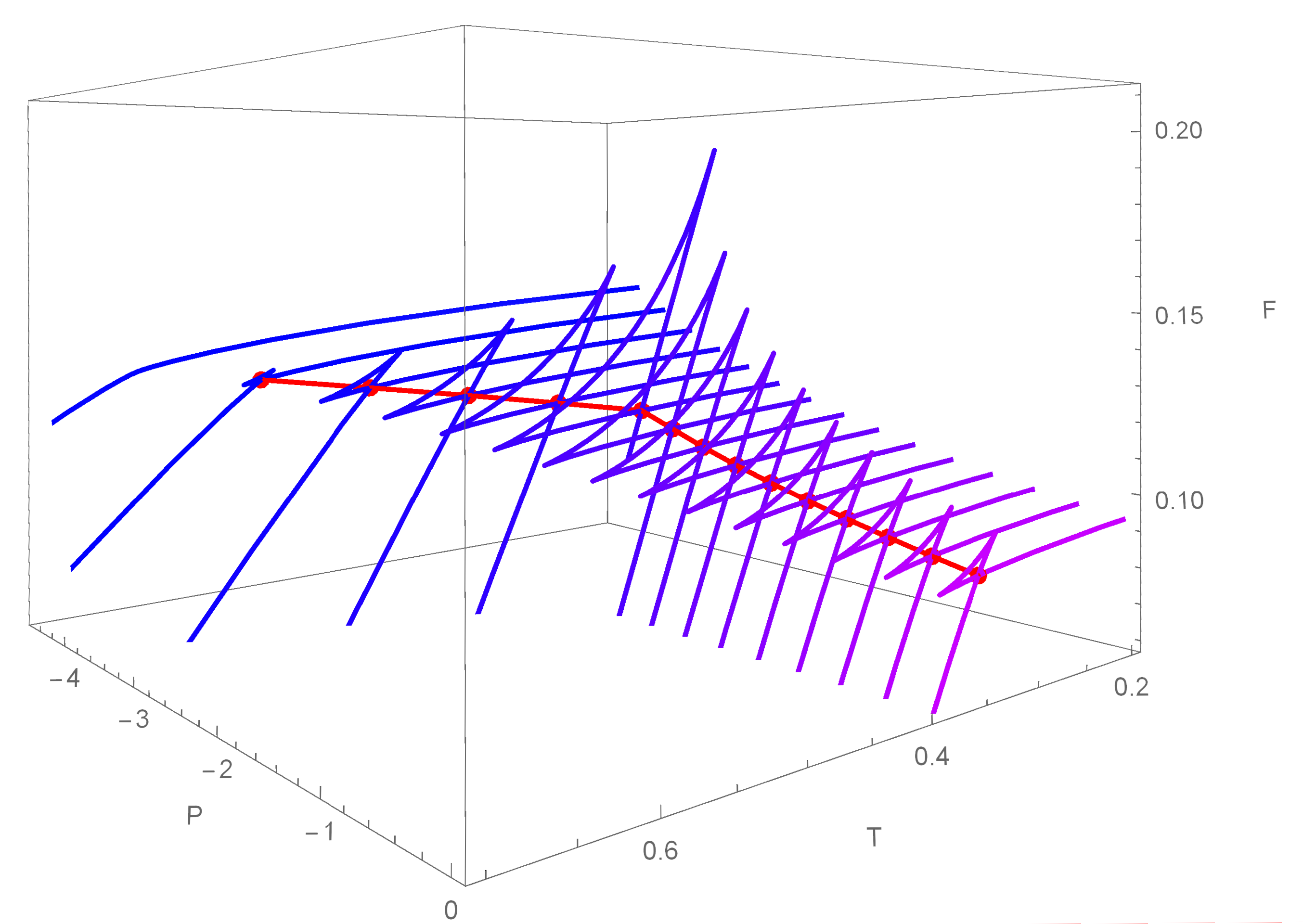}
\caption{ Helmholtz free energy of the Born-Infeld-de Sitter black hole for $q=0.105$, $r_c=0.6$, showing slices of fixed $P$. The red dots mark the small$\rightarrow$large phase transition on each slice. The red line is the coexistence curve. \textbf{Left:} $b=4.6$. A reentrant phase transition is present on the slices indicated with the black arrows. These slices resemble Figure 6a. The remaining slices resemble Figure 6b. \textbf{Right:} $b=5$. There are no reentrant phase transitions present. Note the difference in small-$T$ behaviour between the two figures: on the left, there is a minimum temperature at which the black hole phase exists, while on the right there is no minimum temperature (the lines extend to $T=0$).} 
\end{figure}

Working in the extended phase space also allows us to determine the equation of state for these Born-Infeld-de Sitter black holes. This is a relationship between the pressure $P$, temperature $T$, and volume $V$ (the thermodynamic volume divided by the number of degrees of horizon degrees of freedom) \cite{kubiznak_thermodynamics_2016}.  In AdS space the volume $V\sim r_+^3$ and the specific volume (the volume per horizon degrees of freedom) $v \sim r_+$ \cite{kubiznak_p-v_2012}. However in dS space, \eqref{temp} means that the pressure is a non-linear function of $(T,r_+)$, and \eqref{vol} in turn implies that $r_+$ is a highly non-linear function of $V$. As a result, the equation of state cannot be expressed in closed form. We can plot $P(V)$ at fixed $T$ numerically; however doing so shows an absence of the oscillations characteristic of the van der Waals fluid. We omit the plot here for lack of insight. These oscillations {\it are} present in the Born-Infeld-AdS case examined in \cite{gunasekaran_extended_2012}; it remains to be determined whether the analogy with van der Waals fluids in the extended phase space breaks down because of the presence of the cavity, or due to the fact that we are working in asymptotically de Sitter spacetime. 

\section{Free Energy at Fixed Potential}

As noted above, we should not expect to see a Hawking-Page-like phase transition when examining the free energy as defined by \eqref{F}. This is because the ensemble considered there is one where the total charge $q$ is fixed. However, a black hole cannot dissolve into pure radiation and conserve charge at the same time. Instead, one must consider what happens when the potential is fixed at infinity while the charge is allowed to vary. With these boundary conditions the free energy becomes $G=M-TS-\Phi q$. We examine first the case where $b=\infty$, corresponding to the Maxwell limit, followed by the Born-Infeld case.

\subsection{Maxwell Theory}

In the Maxwell limit, the free energy at fixed potential written as a function of $r_+$, $r_c$, $q$, and $\Lambda$ is:
\begin{align}\label{gibbs1}
G=&\ r_++\frac{q^2}{r_+}-\frac{\Lambda r_+^3}{3}-\frac{\sqrt{3} q^2 (r_c-r_+)}{\sqrt{r_+ (r_+-r_c) \left(r_c r_+ \left(\Lambda \left(r_c^2+r_c r_++r_+^2\right)-3\right)+3 q^2\right)}}\nonumber\\
&+\frac{r_+^3 \left(\left(8 \pi  \sqrt{\frac{r_c}{r_+^2}} \sqrt{(r_c-r_+) \left(r_c r_+ \left(\Lambda \left(r_c^2+r_c r_++r_+^2\right)-3\right)+3 q^2\right)}+\sqrt{3}(\Lambda r_c^2-1)\right)+\frac{\sqrt{3} q^2}{r_c^2}\right)}{4 \sqrt{r_c(r_c-r_+) \left(r_c r_+ \left(\Lambda \left(r_c^2+r_c r_++r_+^2\right)-3\right)+3 q^2\right)}}
\end{align}
\begin{figure}[h]
	\centering
	\includegraphics[width=0.525\textwidth]{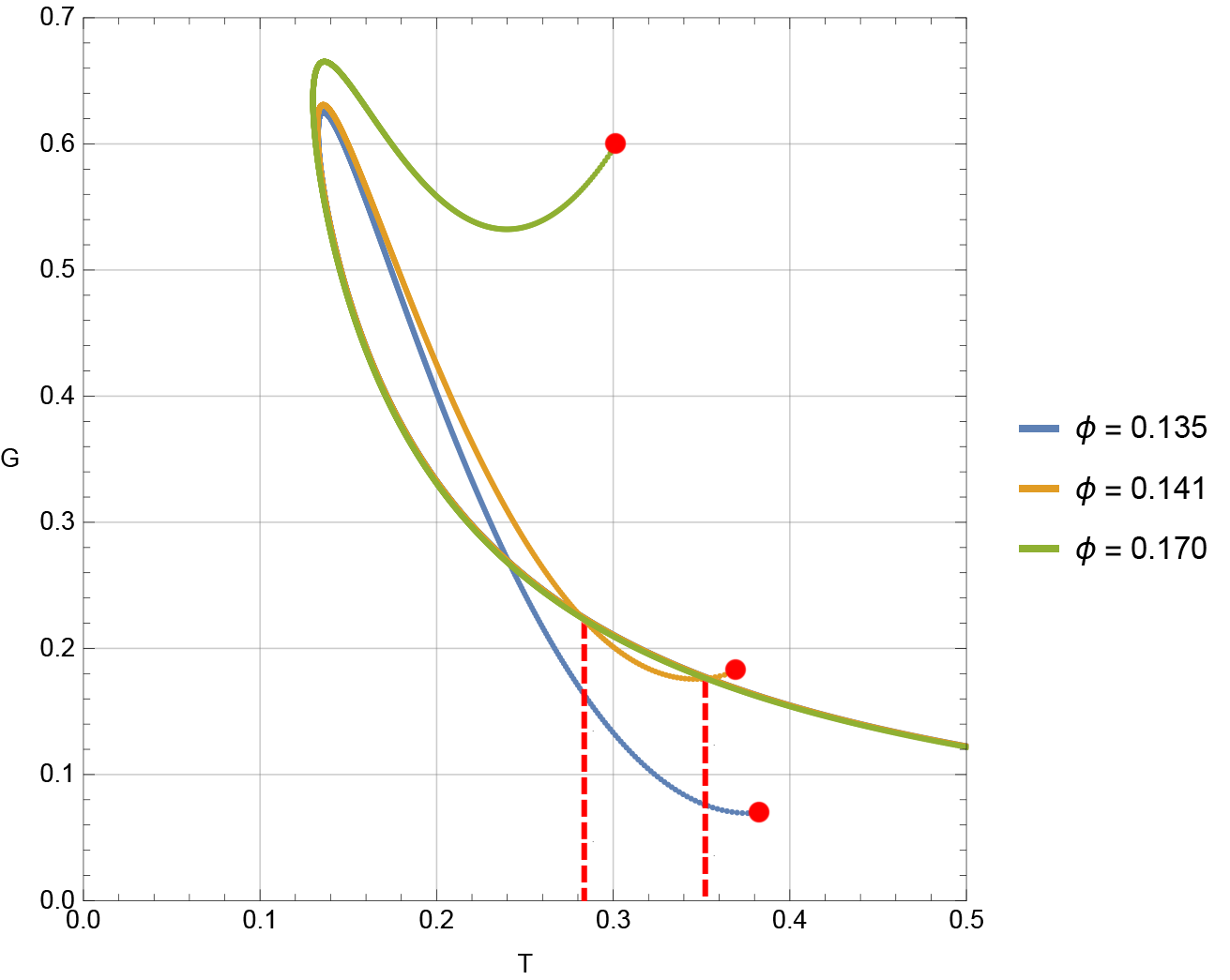}\quad\includegraphics[width=0.44\textwidth]{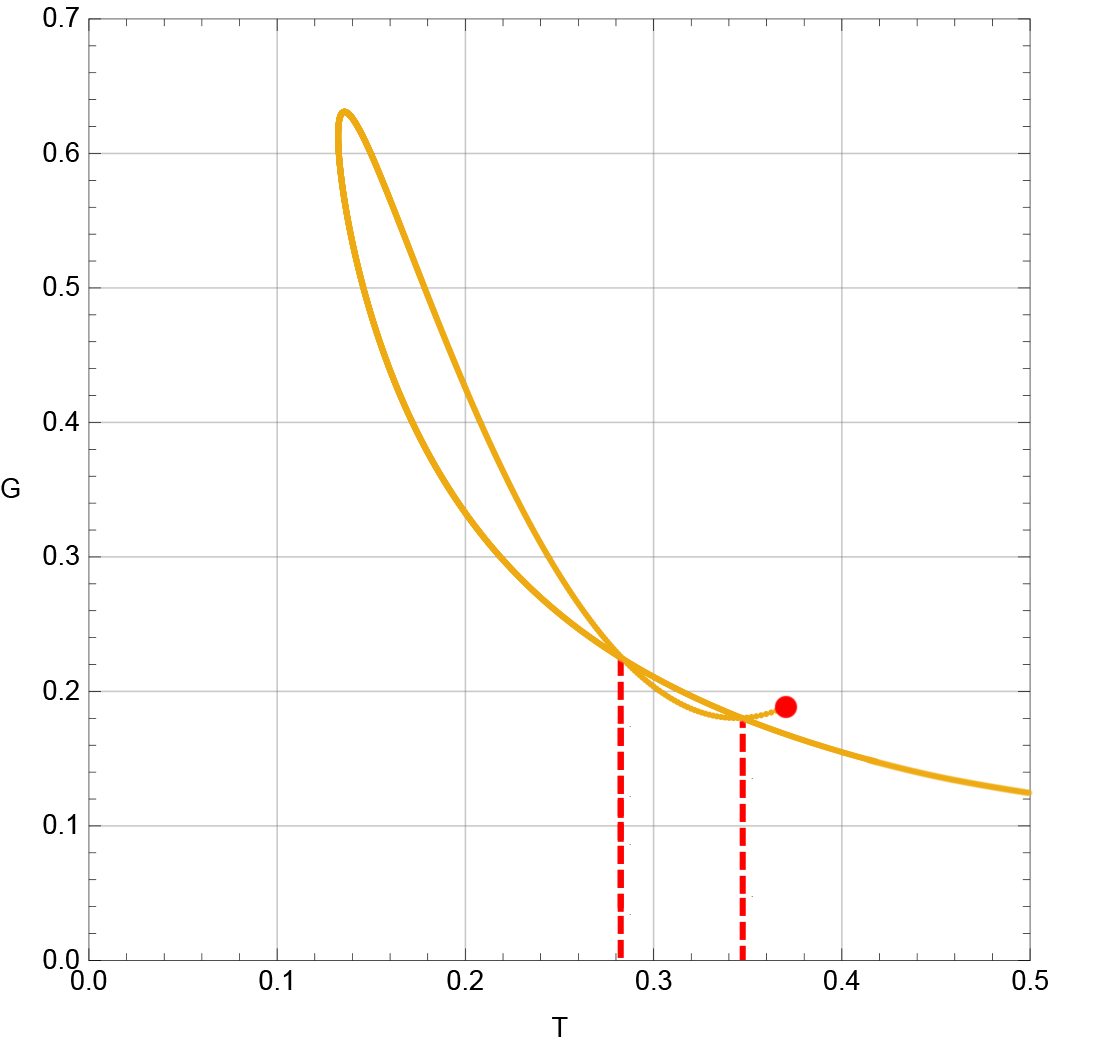}
	\caption{Free energy at fixed potential, with $r_c=1.5$, $\Lambda=0.001$. The temperatures at which phase transitions occur are indicated by the red dashed lines. The cavity radius $r_c$ is indicated by a red dot. \textbf{Left:} The Reissner-Nordstrom-de Sitter black hole, with varying $\phi$. \textbf{Right:} The Born-Infeld-de Sitter black hole with $b=10$ for $\phi=0.141$.} 
\end{figure}

We plot \eqref{gibbs1} as a function of temperature $T$ for fixed $\phi$, $r_c$, $b$, and $\Lambda$ in Figure 8, along with the free energy of the Born-Infeld-de Sitter black hole for large $b$. Examining the figure on the left, as one follows the branches from $T=\infty$ (where they overlap) the black hole increases in size towards the maximal value $r_c$, which is achieved at the red dots. The yellow line in Figure 8a represents a reentrant phase transition, with a small$\rightarrow$large transition occurring at the leftmost red line, followed by a large$\rightarrow$small transition at the second line. For values of $\phi$ above a certain critical value $\phi_c$, the branches no longer intersect (as in the green case), and the small black hole globally minimizes the free energy for all temperatures. This value depends on the choice of $r_c$, $b$, and $\Lambda$ and must be found numerically. Below $\phi_c$ (as in the blue curve) there is only a small$\rightarrow$large transition, as the free energy terminates at $r_c$ before intersecting the small black hole branch again. Critically, these transitions are all {\it metastable}, as the free energy is in all cases above the $G=0$ line corresponding to the radiation phase. \\

\subsection{Born-Infeld Theory}

In the Born-Infeld case, the free energy at fixed potential is:
\begin{align}\label{gibbs2}
G=&\ \frac{4 q^2 \, _2F_1\left(\frac{1}{4},\frac{1}{2},\frac{5}{4},-\frac{q^2}{b^2 r_+^4}\right)-r_+^2 \left(2 b \sqrt{b^2 r_+^4+q^2}-2 b^2 r_+^2+\Lambda r_+^2-3\right)}{3 r_+}\nonumber\\
&-\frac{\sqrt{3\,r^3_+}\,r_c \left(1+r_+^2 \left(2 b^2-\Lambda\right)-2\,b \sqrt{b^2 r_+^4+q^2}\right)}{4\sqrt{r_+}\,X}-\frac{\sqrt{3}\, \textsf{F}}{4  \sqrt{r_+} X} 
\end{align}

\begin{figure}[h]
	\centering
	\includegraphics[width=0.475\textwidth]{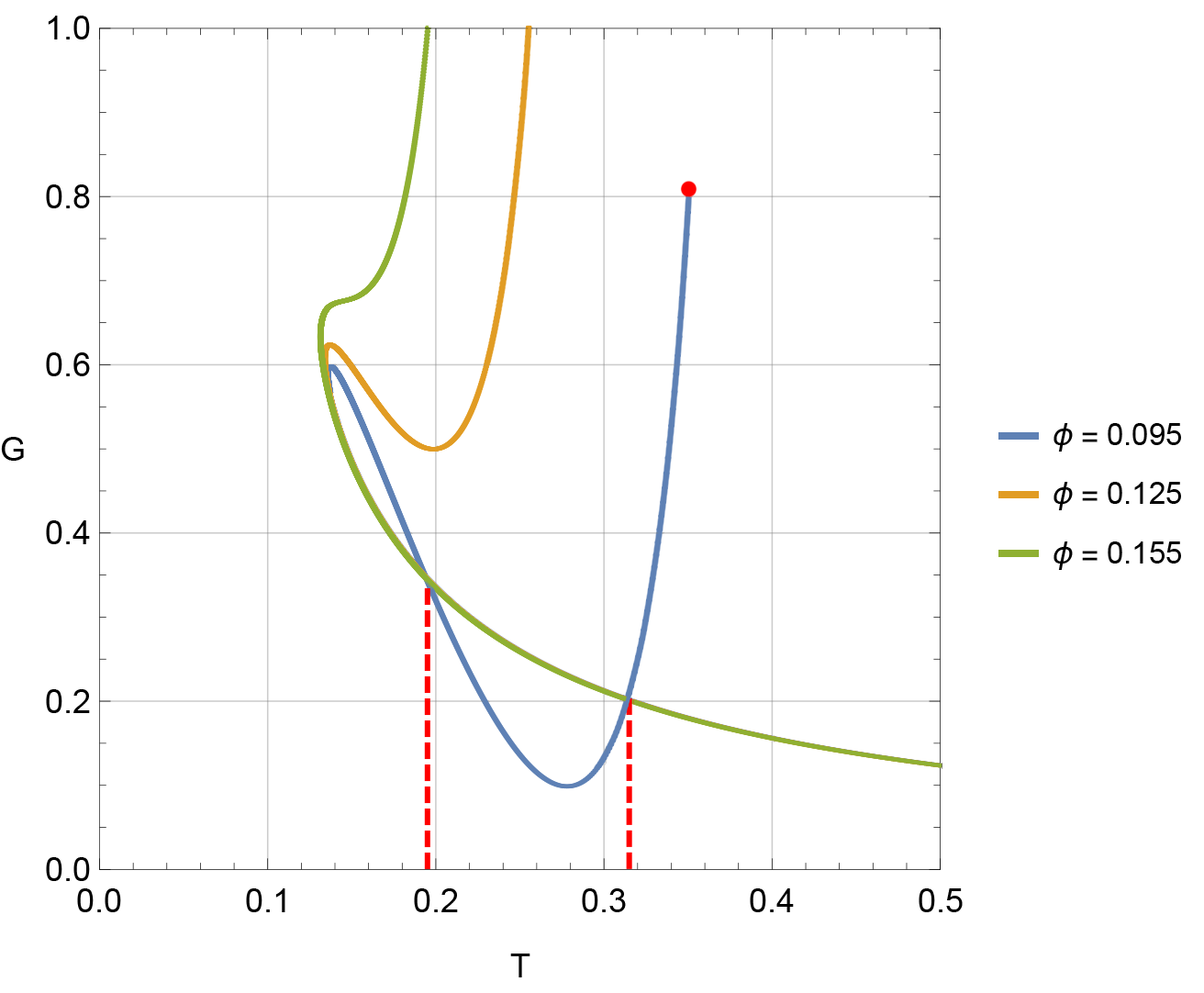}\quad\includegraphics[width=0.485\textwidth]{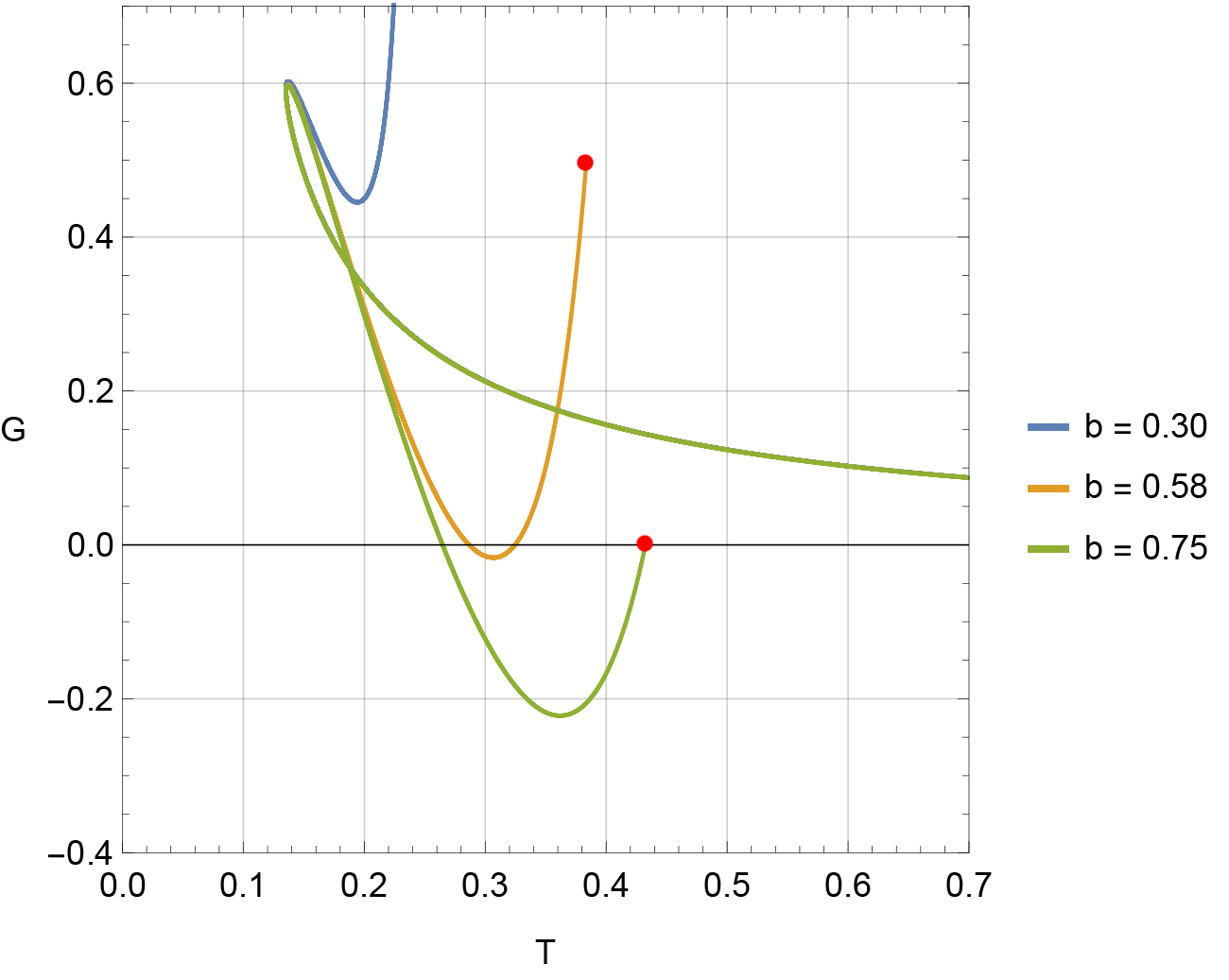}
	\caption{Free energy of the Born-Infeld-de Sitter black hole at fixed potential. \textbf{Left:} $b=0.5$, $r_c=1.5$, and varying $\phi$. The temperatures at which phase transitions occur are indicated by the red dashed lines. The cavity radius $r_c$ is indicated by a red dot. \textbf{Right:} Fixed $\phi=0.095$, $r_c=1.5$, and varying $b$.
	} 
\end{figure}

Figure 9 displays the free energy \eqref{gibbs2} of the Born-Infeld-de Sitter black hole at fixed potential. On the left, we plot $G$ for fixed $b$ and varying $\phi$. Just like in the Maxwell case, there is a critical value of $\phi$ for which the large black hole branch begins to intersect the small branch (as in the blue line) and gives rise to a (metastable) reentrant small$\rightarrow$large$\rightarrow$small phase transition. The exact value of $\phi$ at which this occurs again depends on the value of $r_c$, $\phi$ and $\Lambda$, and must be found numerically. In the given example, this occurs at approximately
$\phi_c\sim0.08$. We emphasize again that the radiation phase globally minimizes the free energy here. For comparison, we plot the same free energy for the Born-Infeld-de Sitter black hole for $b=10$ and $\phi=0.141$ on the right side of Figure 8, demonstrating that for large enough values of $b$ the two cases are identical.\\

On the right of Figure 9, we plot $G$ for fixed $\phi$ and varying $b$. Here we see that a true reentrant phase transition emerges within a small range of values for the maximal electric field strength $b$. In the given example, for $b\in(0.57,0.75)$ the large black hole branch crosses the $G=0$ line, giving rise to a reentrant phase transition from radiation to a n intermediate black hole, and back to radiation as the temperature increases. The thermodynamically stable state is pure radiation at both small and large temperatures,
with a charged black hole as the stable state at intermediate values.
Above $b=0.75$, the cavity radius $r_c$ is reached before the large black hole branch crosses the $G=0$ line, and we have a zeroth order phase transition from a large size black hole to radiation.
To our knowledge this is the first time these particular types of reentrant phase transitions 
from black holes to radiation have been observed. The fact that a black hole exists within an intermediate temperature range while radiation is the preferred state at both low and high temperatures is a curiosity that is so far only seen when an isothermal cavity and Born-Infeld gauge field are present.

\section{Conclusions}

The introduction of an isothermal cavity as an equilibrating mechanism allows for the study of a wealth of thermodynamic phenomena in various asymptotically de Sitter spacetimes. We have shown that Born-Infeld-de Sitter spacetimes admit interesting new phase structures that are not present in either traditional Maxwell theories or in asymptotically AdS spacetimes, namely the existence of a tube-like structure in $F-T-P$ space analogous to the swallowtail typically seen in AdS spacetimes, but which is compact in the sense that there is an upper and lower bound of pressures for which a small/large black hole phase transition occurs. In the Maxwell case studied in \cite{simovic_critical_2018-1}, this {\it swallowtube} represented a series of first-order small$\rightarrow$large black hole phase transitions. The non-linearities of Born-Infeld theory significantly alter this structure, so that each slice of the tube instead represents a {\it reentrant} phase transition  when $b$ is small enough. Notable is the fact that the boundedness of the electric field strength means that the swallowtube is cut off at $P=0$ for a wide range of values for $r_c$ and $q$ (i.e. the phase transition survives in the $P\rightarrow 0$ limit). This is in contrast to the Maxwell case where the tube pinches off at both ends at non-zero pressure, and the phase transition disappears as $P\rightarrow 0$. 
 
We have also observed for the first time reentrant phase transitions between charged black holes
and pure radiation in the Gibbs ensemble where the potential is fixed.  We find that for a certain range of
$\phi$,  Born-Infeld black
holes are stable at intermediate values of the temperature, provided the parameter $b$ is in an intermediate range of values.  Otherwise either cold radiation or hot radiation is the more stable state.

Finally, we note that the isothermal cavity necessarily introduces non-linearities in the equation of state $P(T,V)$ that prevent the oscillations associated with Van der Waals-like phase transitions (typically seen in asymptotically AdS spacetimes) from occurring. This was seen in \cite{simovic_critical_2018-1} in the Einstein-Maxwell case, and remains true even with the inclusion of Born-Infeld electrodynamics. It seems that the equilibrating mechanism (isothermal cavity in dS vs. reflecting boundary conditions in AdS) plays a larger role in determining the nature of the equation of state than does the inclusion of field non-linearity, though it remains to be seen if this is still true in higher dimensions or once angular momentum is considered. Future work will focus on answering this question, as well as mapping out the parameter space further to enlarge our understanding of de Sitter black hole thermodynamics.

\section{Acknowledgements} This work was supported in part by the Natural Sciences and Engineering Council of Canada.

\bibliographystyle{ieeetr}

\end{document}